\documentclass[aps,pre,reprint,groupedaddress,showpacs]{revtex4-1}
\usepackage{graphicx}
\usepackage{dcolumn}
\usepackage{bm}
\usepackage{multirow} 
\usepackage{cancel}

\usepackage{blindtext}
\usepackage{epstopdf} 
\usepackage{amsfonts}
\usepackage{amsmath}
\usepackage{setspace} 
\usepackage[usenames]{color}
\usepackage[dvipsnames]{xcolor}
\usepackage{soul}

\begin{document}
	
	\title{\textbf{\Large Chimera states in a neuronal network under the action of an electric field}}

	\author{Ga\"el R. Simo\textsuperscript{1}, Thierry Njougouo\textsuperscript{1}, R. P. Aristides\textsuperscript{2}, Patrick Louodop\textsuperscript{1}, 
		Robert Tchitnga\textsuperscript{1,3} and Hilda A. Cerdeira\textsuperscript{2,4}\\ 
		\emph{\textsuperscript{1}Research Unit Condensed Matter, Electronics and Signal Processing,
			Departament of Physics, Faculty of Sciences, University of Dschang, PO Box 67, Dschang, Cameroon\\
			\textsuperscript{2}S\~ao Paulo State University (UNESP), Instituto de F\'{i}sica Te\'{o}rica, Rua Dr. 
			Bento Teobaldo Ferraz 271, Bloco II, Barra Funda, 01140-070 S\~ao Paulo, Brazil\\
			\textsuperscript{3}Institute of Surface Chemistry and Catalysis, University of Ulm,
            Albert-Einstein-Allee 47, 89081 Ulm, Germany\\
			\textsuperscript{4}Epistemic, Gomez $\&$ Gomez Ltda. ME, Av. Professor Lineu Prestes 2242,Cietec, 
			Sala 244, 05508-000 S\~ao Paulo, Brazil}}

	\date{\today}
	
	\begin{abstract}
		The phenomenon of the chimera state symbolizes the coexistence of coherent and incoherent sections of a given population. This phenomenon identified in several physical and biological systems
presents several variants, including the multichimera states and the traveling chimera state. Here, we numerically study the influence of a weak external electric field on the dynamics of a network of Hindmarsh-Rose (HR) neurons coupled locally by an electrical interaction and nonlocally by a chemical one. We first focus on the phenomena of traveling chimera states and multicluster oscillating breathers that
appear in the electric field's absence. Then in the field's presence, we highlight the presence of a chimera state, a multichimera state, an alternating chimera state and a multicluster traveling chimera.
		
	\end{abstract}
	
	
	\maketitle

	\section{Introduction}
		Neurons are the building blocks of the brain; these nerve cells are nonlinear elements which communicate with one another between short and long distances \cite{Purves, haken, izhikevich2007}.
		A special connection called the synapse carries such a communication, and there are mainly two general classes of synapses: electrical and chemical. At electrical synapses, the neurons are 
		linked through gap junctions, where channels known as connexons allow the direct flow of electrical currents.  
	On the other hand, at chemical synapses, there is no direct flow of current. Instead, chemical agents called neurotransmitters are carried by 
	synaptic vesicles from one neuron to another \cite{pereda}. And thus, as billion of neurons connect through trillions of synapses  
	to form neuronal networks, they give birth to the brain's complex nature \cite{herrup, swanson}.
	
	Among the complex dynamics that arise from neuronal networks, rhythmic or synchronized behavior between a large number of neurons appears to be key to many neurobiological processes 
	like cortical computation and communication between cortical regions \cite{singer1999,fries,piko}. For instance, it has been suggested that the key event mediating access to consciousness 
	is the early long-distance synchronization of neural assemblies \cite{melloni}. However, synchronization is also related to pathological brain states such as epilepsy, 
	Parkinson's diseases, Alzheimer's, autism, and schizophrenia \cite{uhlhaas}. Moreover, the emergence of synchronized activity microdomains appears to coexist with asynchronized 
	ones on the onset of seizures \cite{jiruska,sejnowski}. 
	As these phenomena are not fully understood, network models to mimic brain rhythms and synchronization patterns are of great interest \cite{rosenblum2017}.
	
	Since the seminal work of Kuramoto and Battogtokh, it is well known that networks of non-locally coupled identical oscillators can split between (synchronized) coherent
	and (asynchronized) incoherent subpopulations \cite{kuramoto2002}. 
	Such spatio-temporal states named \textit{chimera-states} \cite{abrams2004} are known to be stable in the thermodynamic limit. However, for finite-size populations, chimera 
	states have a transient nature and can collapse into the coherent state \cite{wolfrum2011, bick}. Beyond theory, chimera states have been found in neuronal networks, 
	both numerically and experimentally. For example, Santos \textit{et al.} studied chimera-states in a network based on the cat cerebral cortex using the Hindmarsh-Rose neuron model \cite{santos2017,hindmarshr}; Andrzejak
\textit{et al.} studied the link between chimeras and spatiotemporal correlations on intracranial electroencephalography data (EEG) from epilepsy patients	\cite{andrzejak}.
More recently, Sejnowski \textit{et al.} observed chimera states in human electrocorticography data (ECoG) related to seizures onsets \cite{sejnowski}. 
	
	Networks of Hindmarsh-Rose (HR) neurons have been extensively studied on chimera states due to their realistic approximation to neuronal assemblies \cite{storace,majhi}. Hizanidis \textit{et al.} studied 2D and 3D HR neurons for different nonlocal coupling types, finding chimera and mixed
oscillatory states \cite{hizanidis}. 
	Chemical coupling with local, nonlocal, and global interactions was used by Bera \textit{et al.}, confirming the presence of chimera and multichimera states \cite{majhi,bera2016chimera}. Also, nonstationary chimera states
were found in a network of coupled HR neurons with an alternating-current induction through electrical synapses \cite{wei2018}.  However, since the discovery of chimera states, the impact of external electric fields on their existence in neural networks has not yet been investigated. Here we propose to study its effect in a network of Hindmarsh-Rose neurons. And as an external electric field offers the possibility to interact and modify neuronal dynamical states \cite{gluckman96,gluckman01,park05}, we choose the model proposed by Ma \textit{et al.}, which considers the electric field as a new variable coupled to the neuron model \cite{ma2019model}.
	
	 We propose to study the influence of an external electric field on the chimera states in a ring of HR neurons 
	coupled 
	locally under an electrical coupling (gap junctional coupling)  
	and non locally under a chemical coupling. 
	
	We start considering the phenomena that appear in the absence of the electrical coupling and an external electric field. Second, we apply an external electric field on a network with a finite number $N$ of neurons and vary its frequency to study how it influences the chimera states.
Next, we vary $N$ and also different schemes of application of the external field. Last, we analyze the effect of the electrical coupling on the network since both (electrical and chemical coupling) are physiologically important \cite{Michael V.L.,Zahra Shahriari}. In the study, in the absence of an electric field, 
we found new phenomena such as \textit{traveling multicluster chimeras breathers}, {\textit{multicluster chimera breathers,}  which are properly described in Sections III. Among the results obtained when considering an electric field, we can highlight an \textit{alternating chimera state, multichimera states, and a multicluster traveling chimera state} discussed in Section IV.} 
	This paper is organized as follows: in Section II, we introduce the neuronal network under study. Section III explores the chimeras states in the absence of an external electric field while the influence of the number of nearest neighbors in the chemical synaptic coupling is done in Appendix\ref{Ap1}. In Section IV, we study the influence of an external electric field on the network dynamics with and without electrical coupling. Finally, we summarize our results in Section V. In Appendix\ref{Ap2} we apply the \textbf{0-1} test method for the
determination of chaos in some results presented in Section IV\cite{Georg A}.

	\section{Neuronal Network}

	In order to describe our neuronal network, we employ the Hindmarsh-Rose neuron model \cite{hindmarshr} and assume that 
	the neurons are coupled 
	through electrical and chemical synapses. Following the model proposed by Jun Ma \textit{et al.} \cite{ma2019model}, 
	we consider the intrinsic electric field of the neuron, resulting in the following system:
	
	\begin{equation}\label{eq.HR4.mt}
	\begin{cases}
	\dot{x}_i=y_i+ax_i^3-bx_i^2-z_{i}+I+J_i+C_i \\
	\dot{y}_i=1-dx_i^2-y_i+k_1E_i\\
	\dot{z}_i=r\left[s\left(x_i-x_{i0}\right)-z_i\right]\\
	\dot{E}_i=k_2y_i+E_{ext}.
	\end{cases}
	\end{equation}

	The membrane potential of the $i$-th neuron is described by $x_i$, $y_i$ is related to fast currents across the 
	membrane, $z_i$ represents slow currents, and $E_i$ the electric field. $E_{ext}$ and $I$ are the external inputs, while $k_1$ controls the intensity of the electric field effects ($ k_1=0.7 $), and $k_2$ describes the neuron's polarization. It takes the value 0.001($k_2=0.001$).
	The other parameters are set as follow: $a=1$; $b=3$; $d=5$; $r=0.01$; $s=5$; $x_{i0}=-1.6$.
	Our coupling scheme follows the one described by Mishra \textit{et al.} \cite{mishra2017traveling}, where a set of $M$ neurons coupled by electrical and chemical synapses is placed on a ring. The electrical coupling is given by: 
	
	\begin{equation}
	J_i = k_3 \sum_{j=i-1}^{j=i+1}\left(x_{j} - x_i \right) \ .
	\end{equation}

	Here we take into account just the nearest neighbors, and $k_3$ is the electrical coupling strength. 
	The chemical synaptic coupling is written as 
	\begin{equation}
	C_i = \dfrac{k_4}{2p-2}\left(x_s-x_i\right) \Big( \displaystyle\sum_{j=i-p}^{i+p}\Gamma\left(x_j\right) -\displaystyle\sum_{j=i-1}^{i+1}\Gamma\left(x_j\right) \Big),
	\end{equation}
	
	where $k_4$ is the chemical coupling strength, $x_s = 2$ is the reversal potential, which can take values so that the connection turns to inhibitory or excitatory depending on whether $x_s$ is greater or less than $x_i$ \cite{graziane}. Indeed, for specific values of the systems' parameters, the reversal potential is at any time greater than the membrane potential (see Fig.\ref{fig.SGR10}(b)).  For others, there is a reversal of the behaviour of the chemical synapses whose membrane potentials always remain higher than the reversal potential, hence the inhibitory nature (for example see Fig.\ref{fig.SGR2}). {$p$ is the number of neighbors connected on each side except for its two closest ones.} The sigmoidal nonlinear function $\Gamma\left(x_j\right)$ is used to model the chemical 
	synaptic dynamics  \cite{somers,collens}:
	\begin{equation}
	\label{eq.Gamma.mt}
	\Gamma\left(x_j\right)=\dfrac{1}{1+\exp \left[-\lambda(x_j-\theta_s) \right]}.
	\end{equation}
	
	The  parameter $\lambda=10$ determines the slope of the sigmoidal
	function and $ \theta_s=-0.25$ is the synaptic firing threshold. In order to respect the ring topology, we choose periodic boundary conditions, $x_{i}=x_{M+i}$ and the initial condition are given by:
	 $x_i=0.001(i-\dfrac{M}{2})+\zeta_{xi}$, $y_i=0.002(i-\dfrac{M}{2})+\zeta_{yi}$, $z_i=0.003(i-\dfrac{M}{2})+\zeta_{zi}$, 
	where $\zeta_{xi}$, $\zeta_{yi}$, $\zeta_{zi}$ are small random fluctuations. $i=1,...M$.

\section{Phenomena occurring in the absence of an electric field.}
\begin{figure}[!h]
		\includegraphics[width=8.5cm]{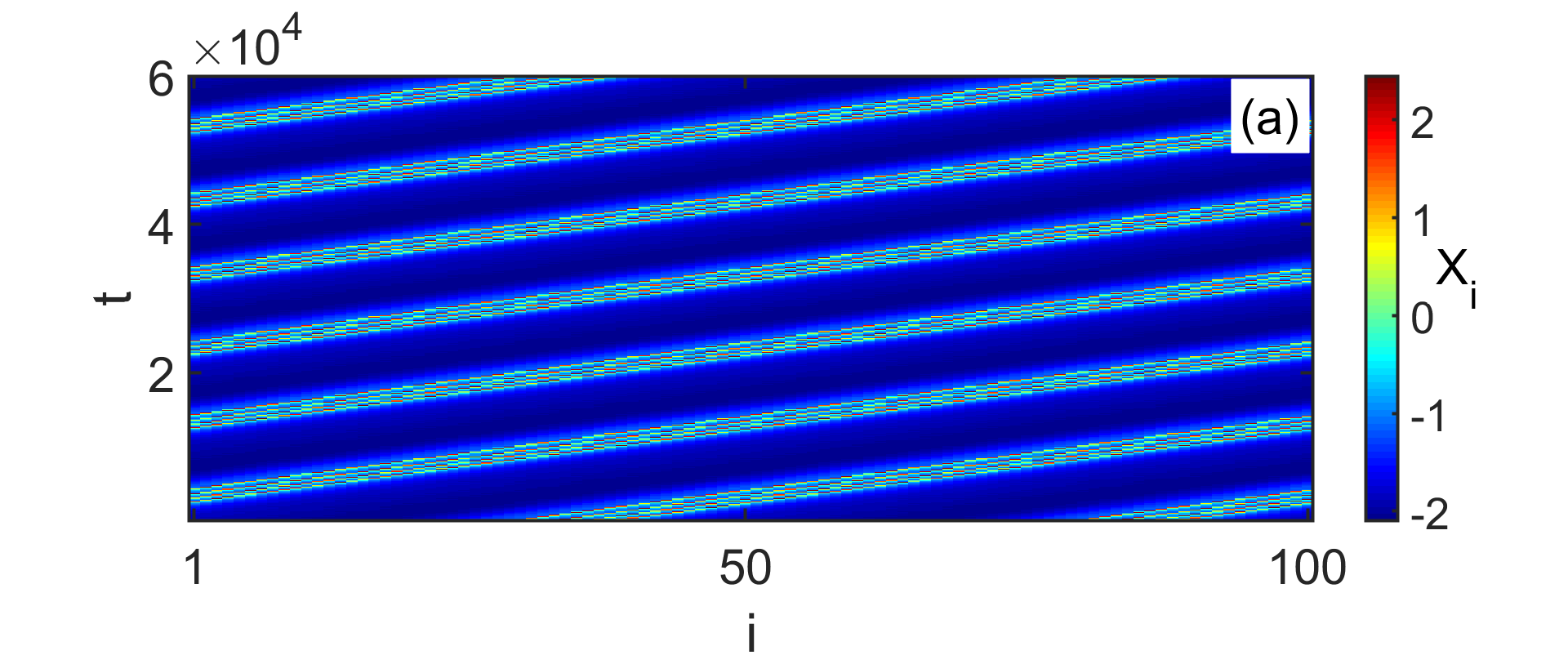}
		\includegraphics[width=8.5cm]{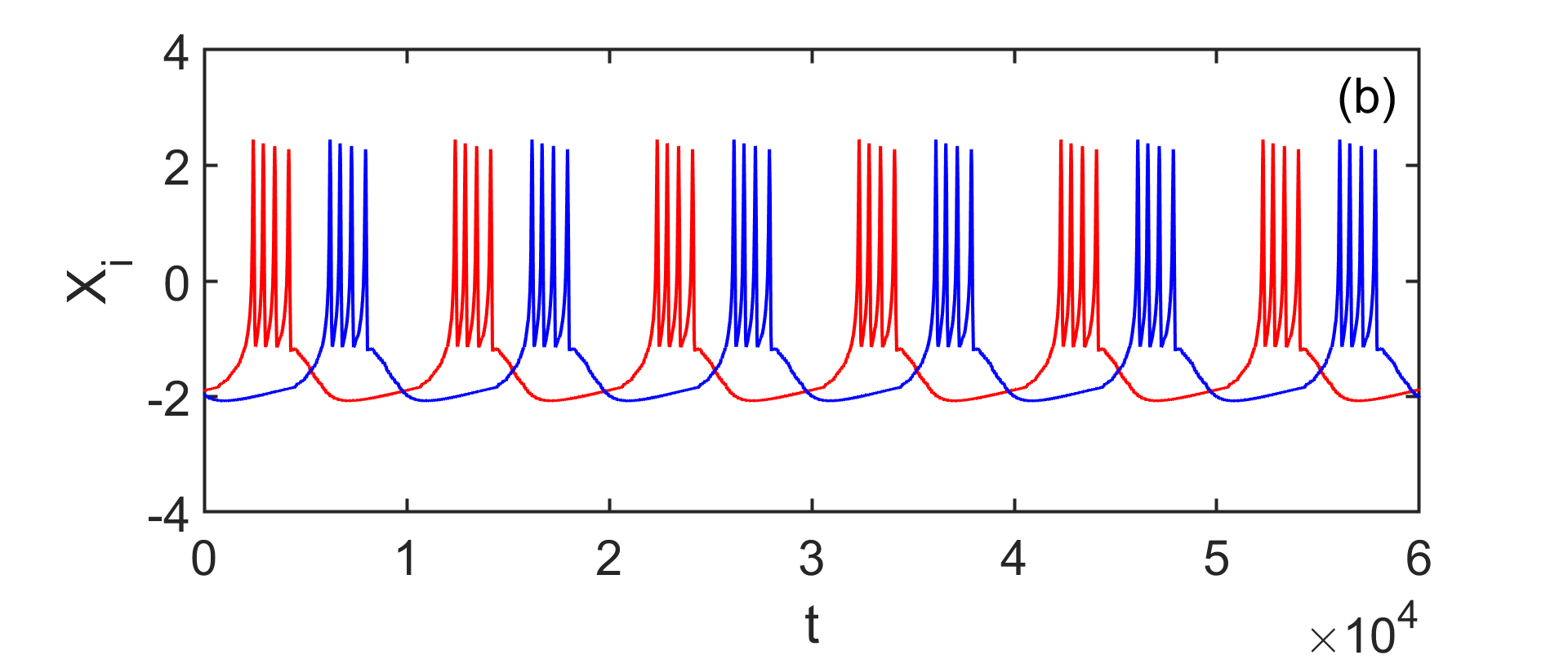}
		\caption{\label{fig.SGR1} Traveling chimera pattern for only chemical synaptic coupling, $k_3=0$, $k_4=9$, $I=3.5$. (a) Spatiotemporal evolution of $x_i$ shows 
			the traveling chimera pattern, where the magnitude of $x_i$ is indicated by the color bar; (b) Time series of $x_i$ of: node 5 in blue and node 50 in red. }
	\end{figure}

	It should be noted that, for relatively small values of the external currents, the network manifests a class of chimera states known as traveling chimera \cite{mishra2017traveling}.  In such states, 
	we see the displacement over time of coherent and incoherent blocks along the ring, where the position of index $1$ is arbitrary. Previous studies \cite{mishra2017traveling} have shown that 
	in the absence of electrical coupling ($k_3=0$), if a chemical coupling strength equal to nine ($k_4=9$) is considered, for a small value of the external current ($I=3.5$) with $p=40$, the elements 
	of the network manifest the traveling chimera while performing each of the regular burstings. This is confirmed by Fig.\ref{fig.SGR1}. By successively assigning larger values to the excitation current,
	we notice that a multicluster is formed for $I=35$, which changes with time, 
	as shown in Fig.\ref{fig.SGR2}. The rest of the work is done with $p = 40$ (See Appendix\ref{Ap1} for more information on the role of $p$).

	\begin{figure}[!h]
		\includegraphics[width=8.5cm]{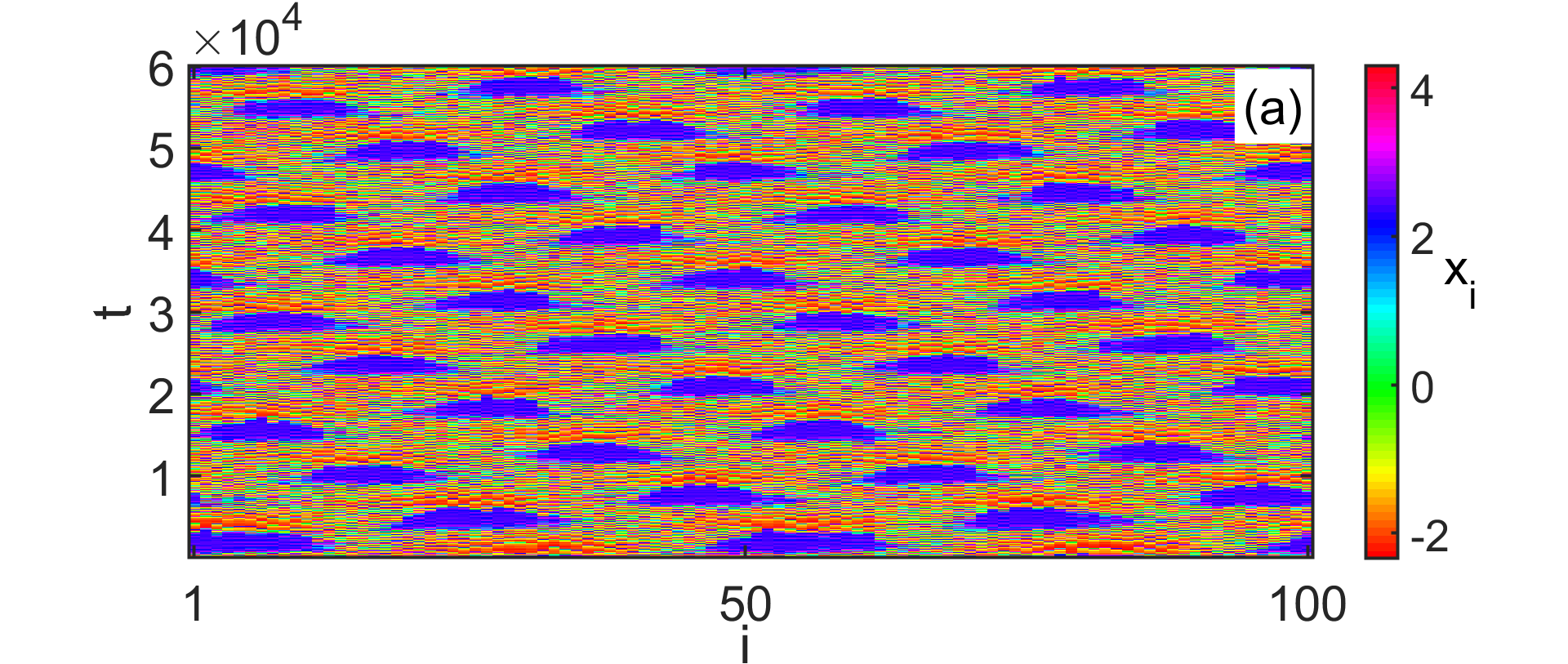}
		
				\includegraphics[width=7.5cm,height=4.5cm]{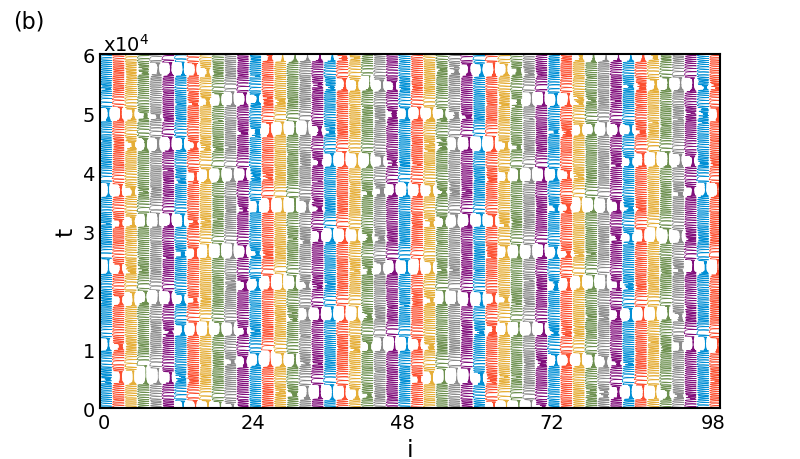}
				
						\includegraphics[width=8.5cm,height=4.5cm]{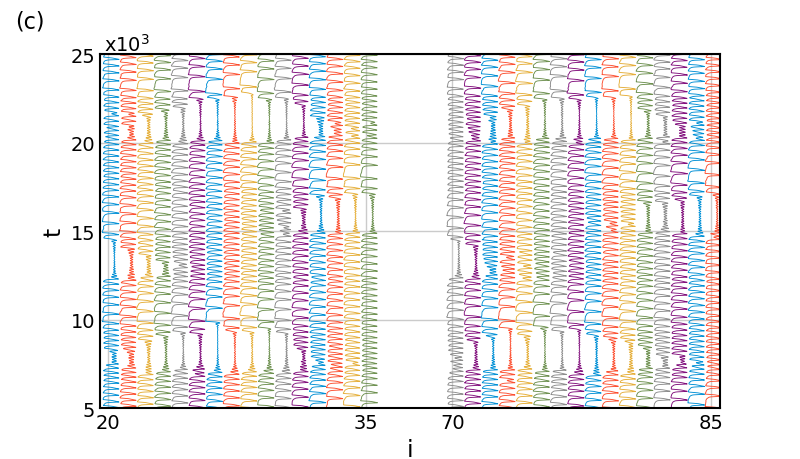}
						
				\includegraphics[width=7.5cm,height=4.5cm]{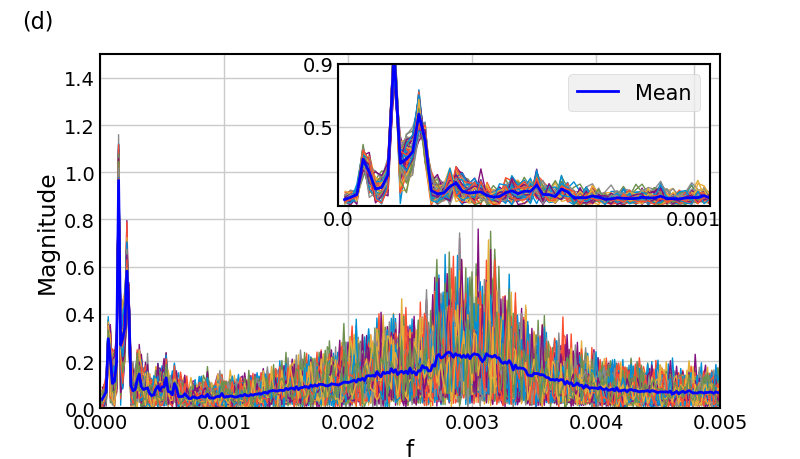}

		\caption{\label{fig.SGR2} 
		 Traveling multicluster for $k_3=0$, $k_4=9$ and $I = 35$. (a) Spatiotemporal evolution of the $x_i$. Coherent clusters in blue. 
			(b) Plot of time series of neurons with even index $i$ for better visualization of coherent clusters. 
			(c) Close up on figure (b), where the coexistence of coherent regions, marked by low-frequency behavior, is  distinctly seen.
			(d) Fourier transform for all oscillators in the ring, with low frequencies expanded in the inset. Notice that main low frequencies are common to all. 
			}
	\end{figure}
	
 The spatio-temporal evolution of the variable $x$ of each of the $M$ neurons of the network is given in Fig. \ref{fig.SGR2}(a), where there appear two large subpopulations, 
 the first one in blue which forms two large blocks of coherent neurons and the second subpopulation of incoherent neurons in multicolor (with a tendency to form a second 
 coherence region in red color)
interspersing the coherent regions. 
 In order to clarify the behavior of the system, we plot the time series of all oscillators in such a way that their time behavior becomes apparent (see Fig.\ref{fig.SGR2}(b) and (c)). Putting together these two graphical representations of the same phenomena  permits us to distinguish the  coherence regions from the incoherence. We see in Fig.\ref{fig.SGR2}(b) that the coherent regions (blue in \ref{fig.SGR2}(a)) correspond to the low-frequency regions in the neuron's  behavior, and this allows us to follow the clusters in Fig.\ref{fig.SGR2}(b) because they correspond to the white regions. We may also notice that more than one coherent regions exist simultaneously (see Fig.\ref{fig.SGR2}(c), where we have eliminated oscillators to facilitate interpretation). 
This coexistence of several coherent subpopulations alternating with incoherent ones in the network is identified in the literature as a multicluster chimera state \cite{yao2015emergence,travchime}. These coherent blocks do not keep a fixed position in time and space. We note here a shift from the bottom right corner to the upper left corner of coherent neuronal blocks along the ring in time, which is a manifestation of a phenomenon reported here for the first time, to the best of our knowledge. A Fourier analysis of the time series of all oscillators shows some low frequencies common to all. Therefore these traveling multiclusters are indeed periodic, which give rise to \textit{traveling multicluster chimera breathers}, since the coherent blocks appear and disappear periodically (see Fig.\ref{fig.SGR2}(a) to (c)). Fig.\ref{fig.SGR2}(d) which shows the superposition of the Fourier transform for all oscillators in the ring, corroborates that there exist several periodic slow motions in the system which can be observed in Fig.\ref{fig.SGR2}(a). \\

	\begin{figure}
		\includegraphics[width=9.5cm]{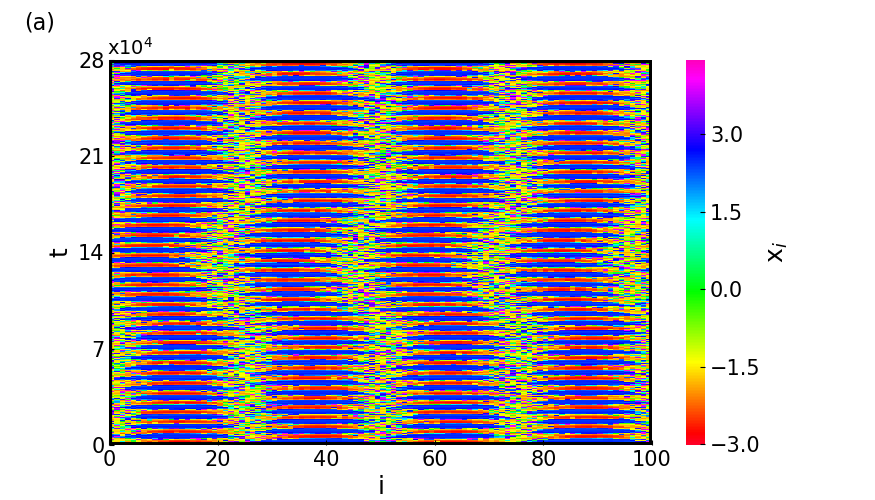}
		
				\includegraphics[width=8.5cm]{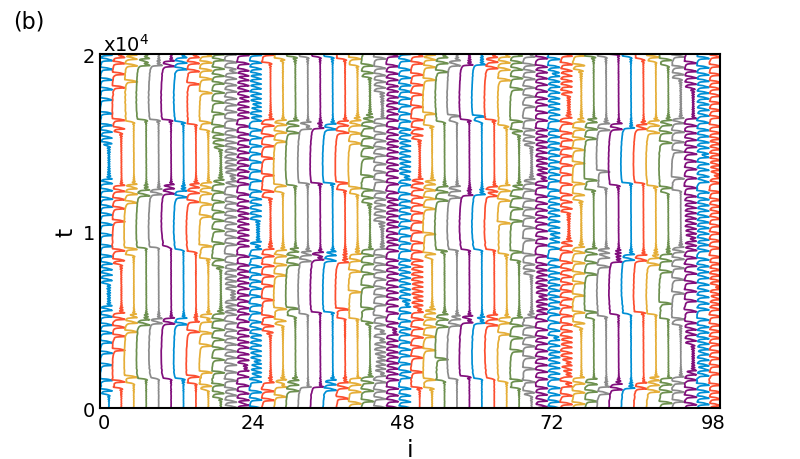}
		
				\includegraphics[width=8.5cm]{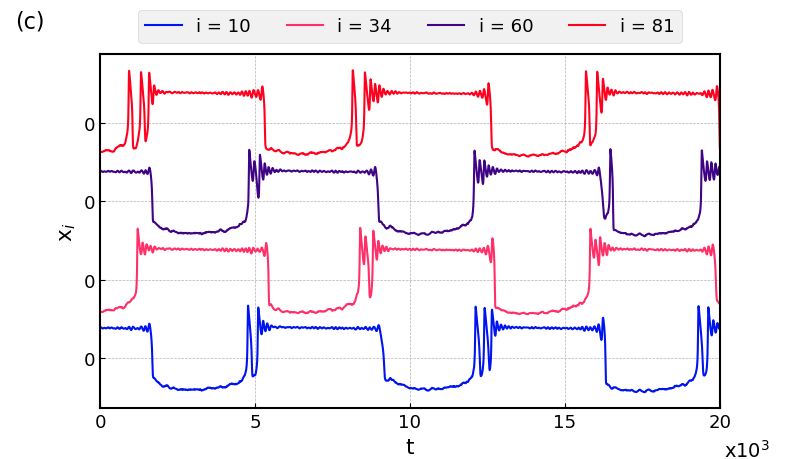}
				
						\includegraphics[width=8.5cm]{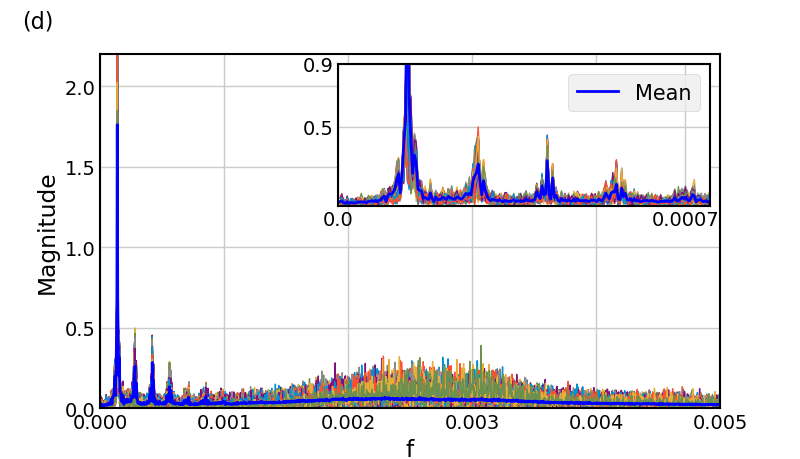}

		\caption{\label{fig.SGR3} 
			Multicluster chimera breathers, for $k_3=0$, $k_4=10$, $I=35$. (a) Spatiotemporal evolution of $x_i$, marked by four coherent regions. (b) Close-up of (a), details of $x_i$ show the existence of multiclusters.
		 (c) Time-series of $x_i$ for neurons inside different but subsequent clusters show that these present an ``antiphase'' like behavior.
            (d) Superposition of Fourier transform of all oscillators in the ring. Very well defined low frequencies, which are shown in the inset, are common to all. }

	\end{figure}

Increasing the chemical coupling strength $(k_4 = 10)$ results in a different spatio-temporal evolution; this is marked in Fig. \ref{fig.SGR3}(a) by the presence of four  coherence clusters, spatially separated by incoherent domains. Firstly, we can divide the coherent clusters into two domains, and at each instant of time, we see that the system presents two larger blue clusters (where $x_i>0$) and two smaller red ones (where $x_i < 0$). Moreover, we note that the size of each cluster oscillates in time, which is  manifest in Fig. \ref{fig.SGR3}(b). This oscillatory behavior has also been observed in a chimera state with two clusters \cite{zhu2012}.
If we assume that the blocks in Fig. \ref{fig.SGR3}(a) are numbered from $1$ to $4$, starting to the left, we see in Fig. 3 (c) that neurons inside block $1$ are not only coherent with neurons inside cluster $1$, 
but are also coherent with neurons of cluster $3$. The same behavior occurs with neurons from clusters 2 and 4. 
On the other hand, neurons from blocks 1 and 3 are in antiphase with neurons from clusters 2 and 4. A similar result was reported in \cite{laing2009} but with spatially stationary clusters. Well-defined lower frequency peaks mark the Fourier analysis for this case, as seen in Fig. \ref{fig.SGR3}(d). Here we see that all oscillators share those low frequeny motions which again as in the case of Fig. \ref{fig.SGR2}, can be seen from Fig. \ref{fig.SGR3}(a).

Another feature of this chimera state is the motion along the ring. As seen in Fig. \ref{fig.SGR3}(a), the location of clusters varies in time as they oscillate in size while drifting to the left. Putting all these features together, we conclude that we have found for the first time a \textit{multicluster chimera state} whose size varies periodically in time and space (breathers) 
Zhu et al.\cite{zhu2012} observed a similar effect with
only two coherent clusters, which they named \textit{two-cluster oscillating chimera state}. Since our system has more clusters that synchronize at different values while oscillating in time and space, we would call them \textit{Multicluster chimera breathers}.

	\begin{figure}
		\includegraphics[width=8.5cm]{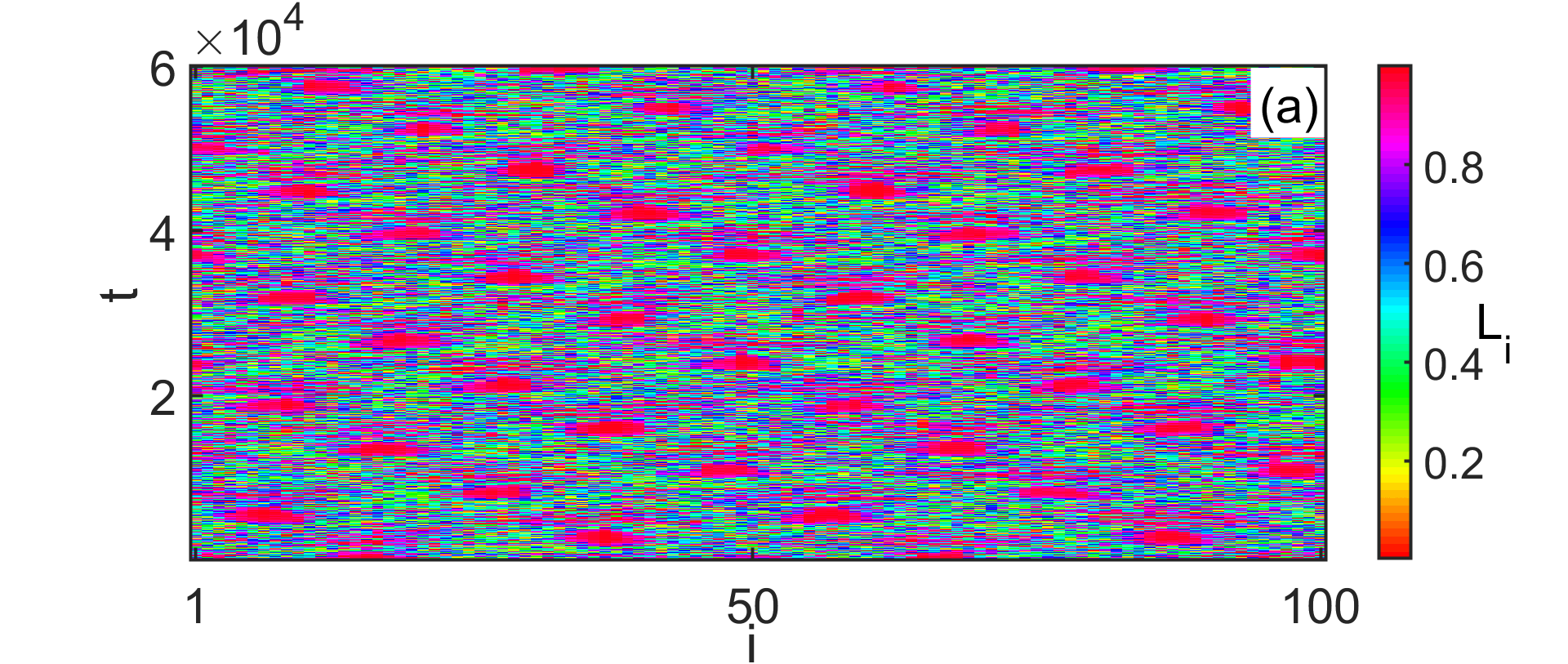}
		
		\includegraphics[width=8.5cm]{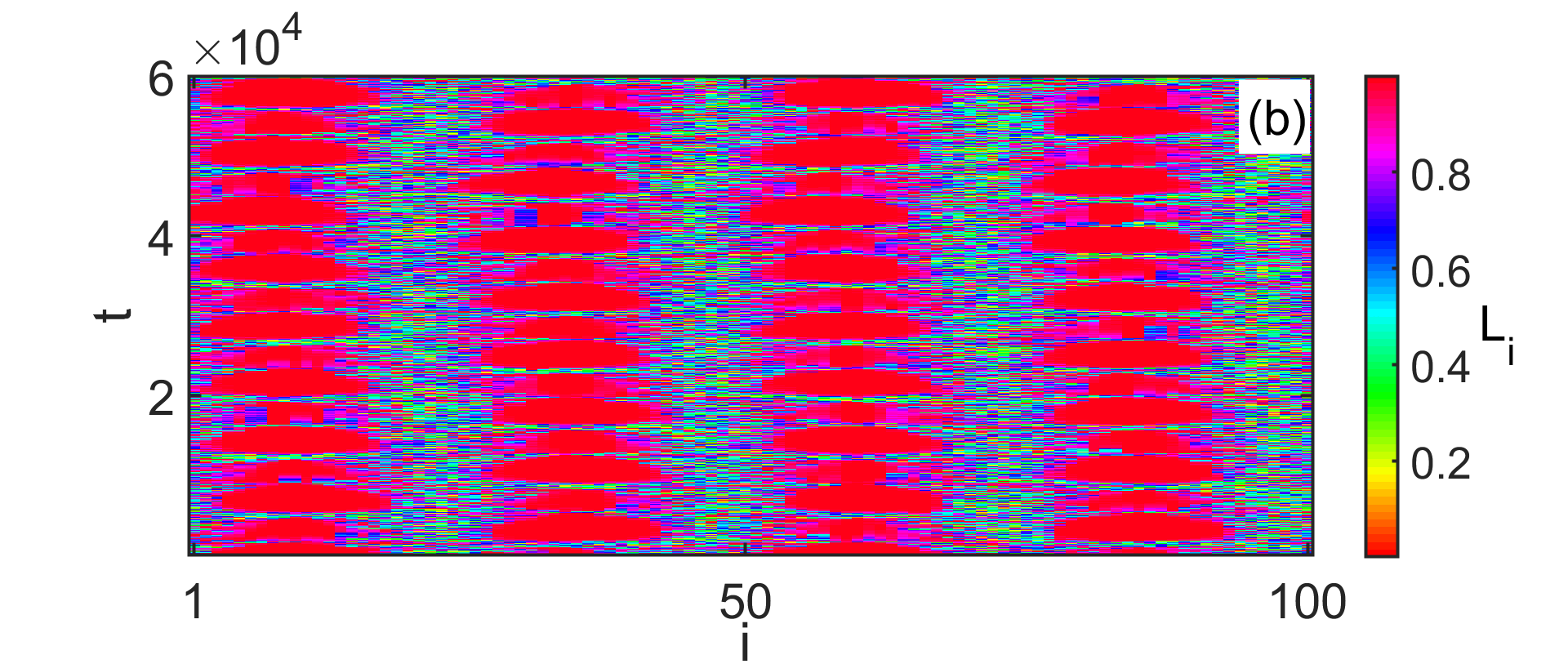}
		
		\caption{\label{fig.SGR4} 
			Spatiotemporal evolution of the local order parameter $L_i$ with $\eta = 2$, the magnitude of $L_i$  is indicated using color bars: (a) For the parameters of Fig.\ref{fig.SGR2}; (b) For the parameters of Fig.\ref{fig.SGR3}.}
	\end{figure}

	To confirm the state of coherence of the blocks of neurons, we called upon the notion of the local order parameter \cite{dai2017two,omelchenko2011} 
	defined by: \begin{equation}\label{eq.Li.mt}L_i=\left|\dfrac{1}{2\eta}\sum_{\left|i-k\right|\leq{\eta}}\mathrm{e}^{\mathrm{j}\Phi_k}\right|,\end{equation} 
	where $\eta$ represents the number of neighbors to the left and right of neuron $i$, $j=\sqrt{-1}$ and $\Phi_k$ the geometric phase is given by the expression: 
	\begin{equation}\label{eq.Phi.mt}\Phi_k=\arctan{(\dfrac{y_k}{x_k}).}\end{equation}

As we calculated $L_i$ for all neurons in the ring, we can differentiate coherent states, where $L_i$ is close to unity, to regions of incoherence, where $L_i$ decreases. Figure \ref{fig.SGR4} confirms the previous statements. First, in Fig.\ref{fig.SGR4}(a), we present the spatio-temporal evolution of $L_i$ for ($k_4 = 9$), where we see not only that the position of coherent clusters varies with time, but they also present the breathing behavior marked by intervals where the system is incoherent. Second, in Fig.\ref{fig.SGR4}(b), where ($k_4 = 10$), we note the four distinct clusters and how they vary in size within time.  
	
	\section{Application of the Electric Field}
	
	In this section, we consider when a weak electric field is
applied to parts of the network under study. The aim is
to study its impact on the states previously presented and
use it for a possible control. For the sake of simplicity, we choose the following form of the external electric field:
	\begin{equation}\label{eq.Eex.mt}E_{ext}=E_{m}\sin(2{\pi}ft),\end{equation} 
	where $E_{m}$ is the amplitude of the field and $f$ its frequency.
	To start with, we neglect the electrical coupling ($k_3=0$) and work with a relatively small value of the excitation current.
	Keeping the value of the chemical coupling strength at {$k_4=9$} as before, and  $I=3.5$, we apply the electric field ($k_1\neq0$), with $E_m = 1.5$, to the last 50 oscillators in the ring ($N=50$), where we are following the same indexing as before, section III. 

\begin{figure}
	\includegraphics[width=8.5cm]{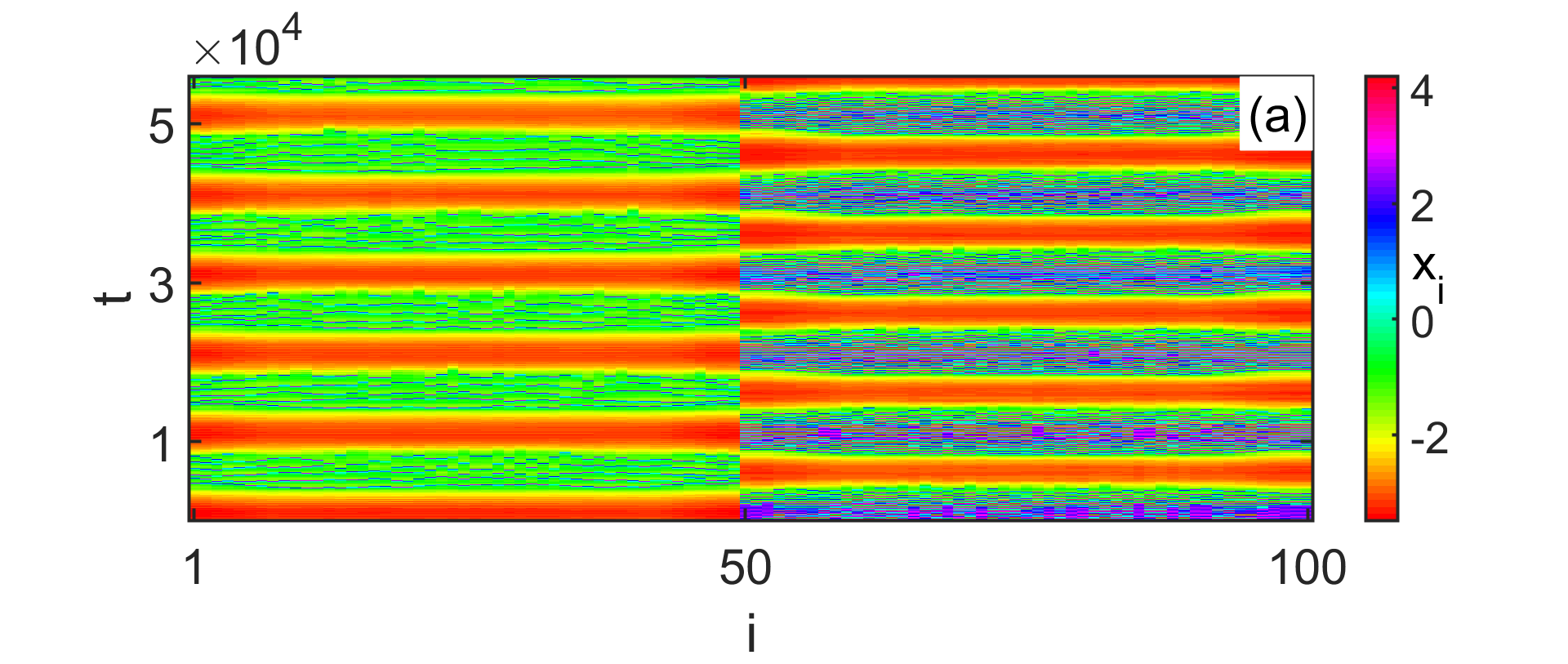}	
	\includegraphics[width=8.5cm]{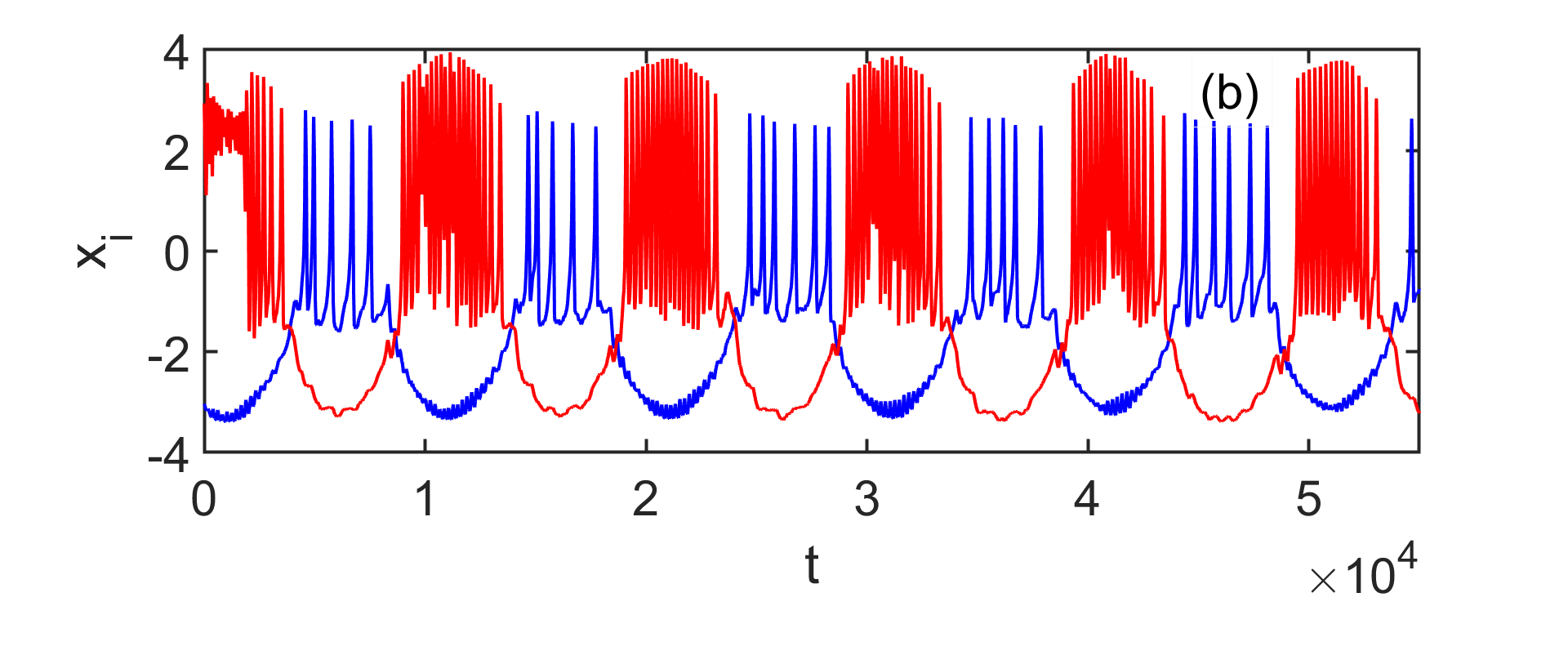}
	
	\includegraphics[width=8.5cm]{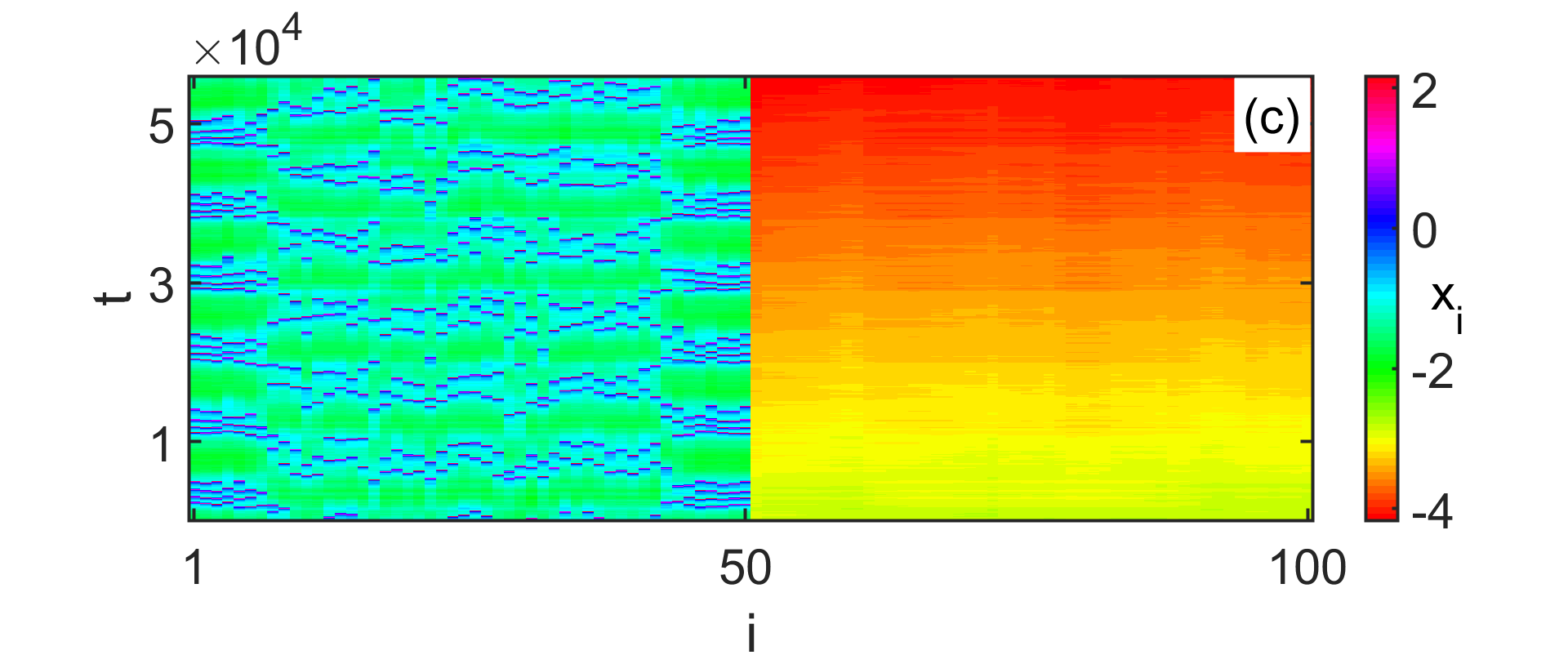}
	\includegraphics[width=8.5cm]{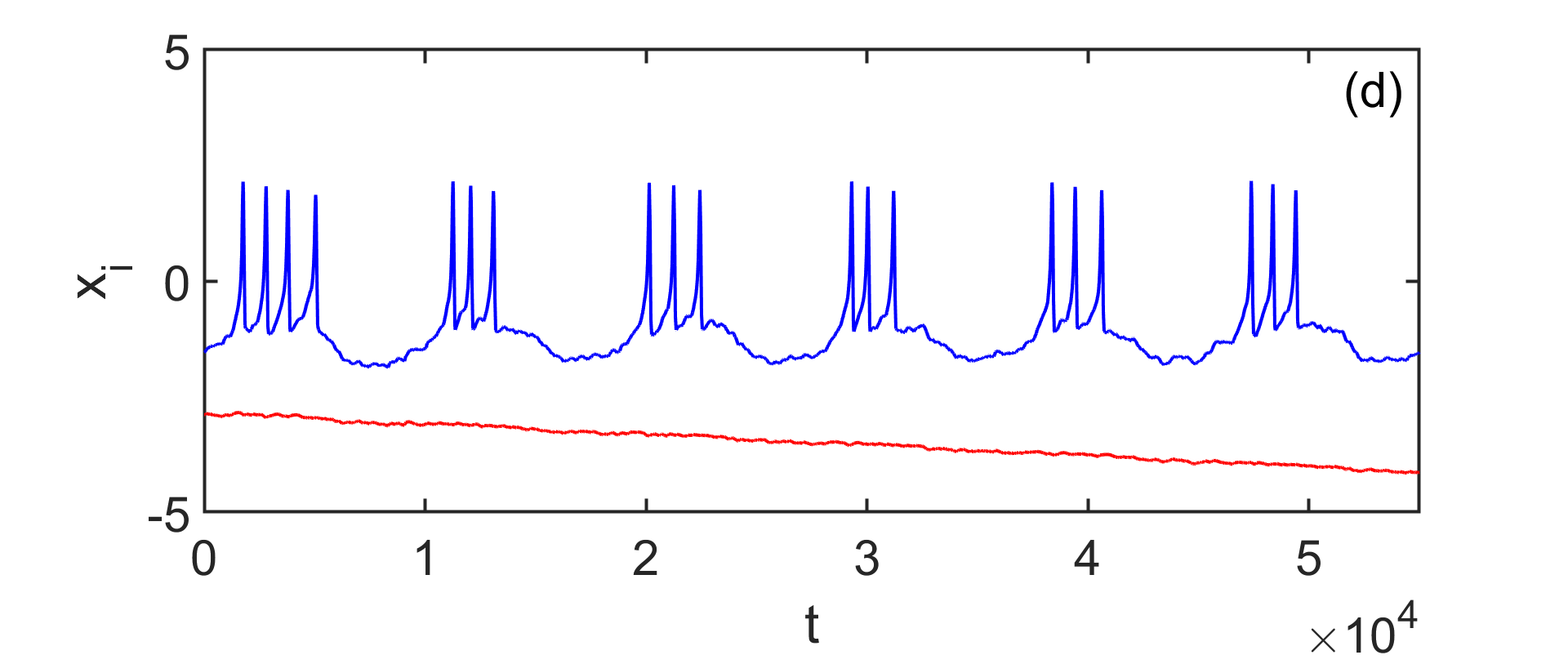}

	\caption{\label{fig.SGR5} 
		Influence of frequency $f$: For $f=0.01$: (a) Spatio-temporal evolution of the $x$-variables; (b) time series of the $x$-variables for 
		neurons in the immersed(red) and non-immersed (blue) regions, where color bars show the amplitude values of the x-variables, 
		For $f=12$: (c) Spatio-temporal evolution of the $x$-variables; (d) time series of the $x$-variables for 
		neurons in the submerged (red) and non-submerged (blue) regions. Color bars show the amplitude values of x-variables.}

\end{figure}

 Varying the frequency of this field induces many different situations. For $f = 0.01$, the electric field induces a chimera state known as alternating  chimera \cite{haugland}, characterized by an exchange of dynamics between regions of the network. As (Fig. \ref{fig.SGR5}(a)) shows,  we have a coherent cluster formed by the first $50$ neurons in the initial state. The rest of the network, where neurons are immersed in the electric field, is incoherent. However, after some time, the initially coherent region becomes incoherent while the incoherent becomes coherent. We also note that the period between bursts seems to be independent of the electric field,(Fig. \ref{fig.SGR5}(b)). 
 We checked values of $f$ greater than 0.01 and smaller than 200 and noticed that 
	there appears a chimera state characterized by the presence of incoherent zones for those oscillators where the electric field has not been applied and a coherent zone coinciding with the part subjected to the electric field. This effect we show in (Fig. \ref{fig.SGR5}(c)) for $f=12$. 
	The temporal evolution is totally different in the two regions: the neurons are bursting in the nonsubject part, and they are not bursting when the part is subjected to the field (Fig. \ref{fig.SGR5}(d)).

\par

We are interested now in the network's chimera state for the interval of frequencies mentioned previously ($f>0.01$), particularly in the case of $f=12$. 
To quantify the chimera state in the network, we calculate the strength of incoherence ($SI$), a statistical metric 
recently introduced by Gopal \textit{et al.} \cite{gopal2014observation}, which can characterize the incoherent, chimera, and coherent states in arrays of oscillators. 
So if we consider an array of $M$ oscillators, from which a chimera state emerges in the permanent regime, we first find the asymptotic state variables $zz_i$, defined as the difference between the state variable of two nearest neighbors, i.e., $zz_i=x_i-x_{i+1}$. 
The estimate of $SI$ is based on the local standard deviation of $zz_i$ computed in $N_b$ bins of equal length $n=M/N_b$:

\begin{equation}\label{eq.sigma.mt}
\sigma(m;t)=\sqrt{\frac{1}{n} \sum_{j=n(m-1)+1}^{mn}\left[ zz_{j}-\langle zz_{m}\rangle\right]^2}, 
\end{equation}
where $m=1,...,N_b$.
Then, the  local standard deviation of the asymptotic state is defined as the time average of 
Eq. (\ref{eq.sigma.mt}): \begin{equation}\label{eq.sigma1.mt}\sigma(m)=\langle\sigma(m;t)\rangle_t.\end{equation} 
A threshold $\delta$ is introduced, and one defines $s_m=\Theta(\delta -\sigma(m))$, 
where $\Theta(\cdot)$ is the Heaviside function. Hence, if the fluctuations in the $m^{th}$ bin 
are less (greater) than $\delta$, a flat (sparse) segment is found, and  $s_m=1$ ($s_m=0$). 
Hence, $s_m$ discretizes $\sigma(m)$ and provides a symbolic coding for the coherent and incoherent segments.
Finally, the strength of incoherence (for the asymptotic state) is estimated 
as:\begin{equation}\label{eq.SI.mt}SI=1-\frac{1}{N_b}\sum_{m=1}^{N_b}s_m,\end{equation}
being $0$ for a coherent state, $1$ for an incoherent one, and $0<SI<1$ for chimera 
(and multichimera) states.

  For the case of Fig. \ref{fig.SGR5}(c), we have $SI = 0.666$, confirming that it is a 
  chimera state. To better understand and distinguish further between chimera 
  and multichimera states,  we calculate \cite{gopal2014observation}
  the discontinuity measure $DM$ based on the distribution of $s_m$ along the ring 
  defined as
  \begin{equation}
  \label{eq.DM.mt}
  DM=\frac{\sum_{m=1}^{N_b}\left|s_m-s_{m+1}\right|}{2} \ . 
  \end{equation}
It is specified that the discontinuity measurement takes the value 0 
  when there is total coherence or total decoherence; it takes 1 for a chimera state with a coherence domain and integer 
  values greater than 1 and less than $M/2$ for multichimera states \cite{gopal2014observation}.

\begin{figure}[!h]	
	\includegraphics[width=7.5cm]{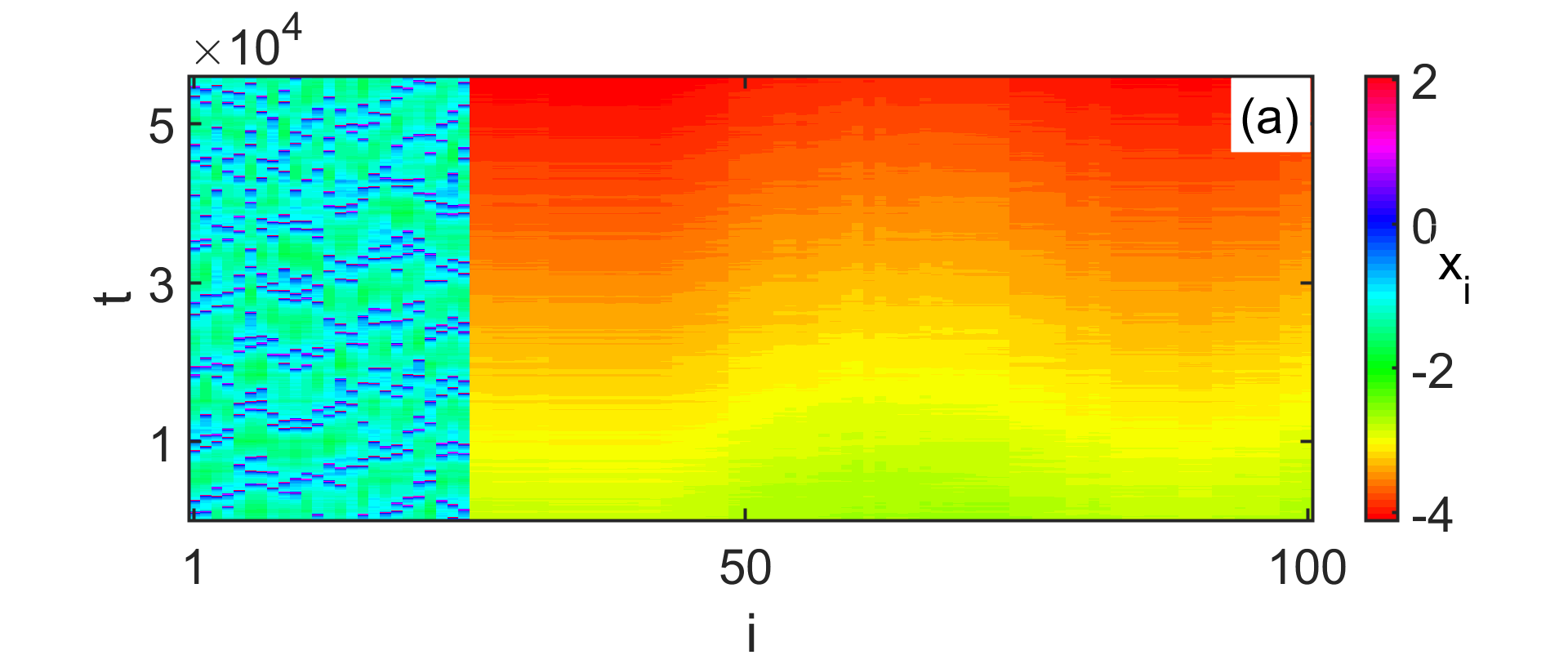}	
	\includegraphics[width=7.5cm]{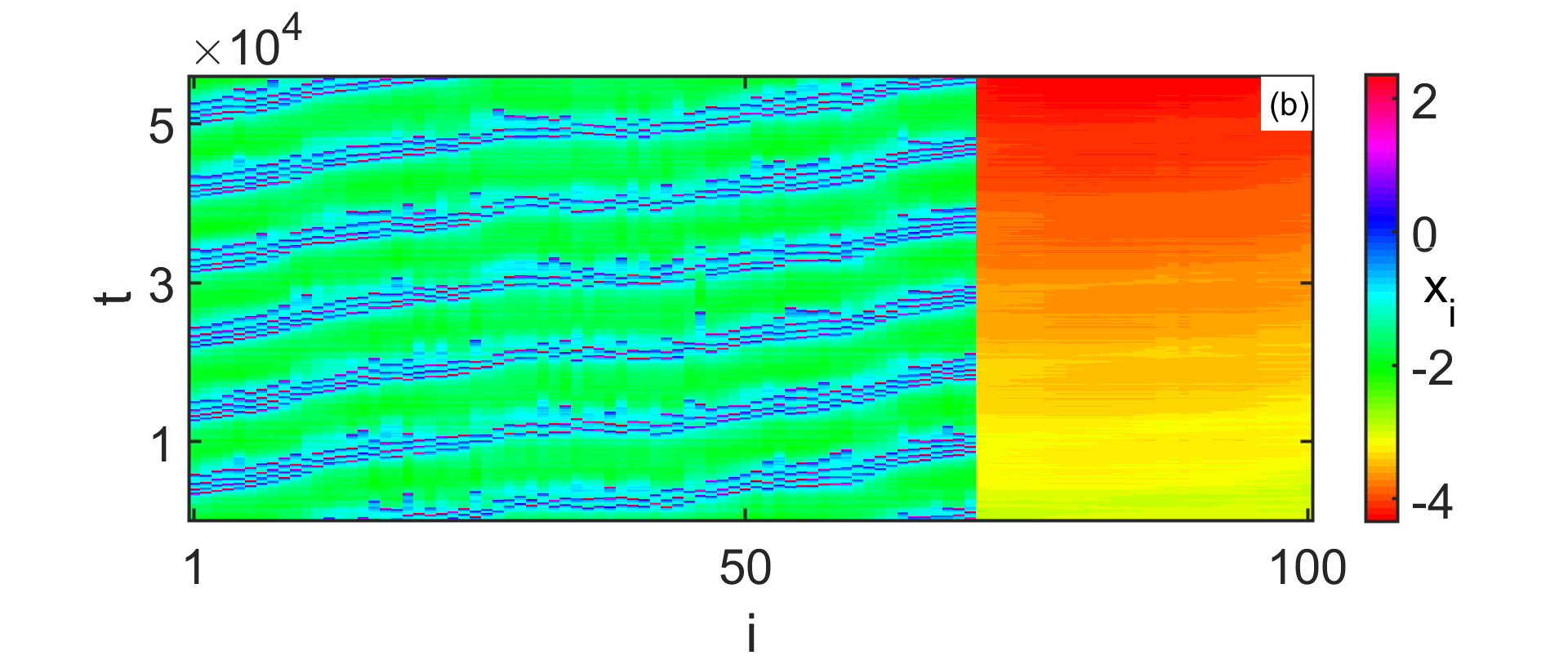}	
	\includegraphics[width=7.5cm]{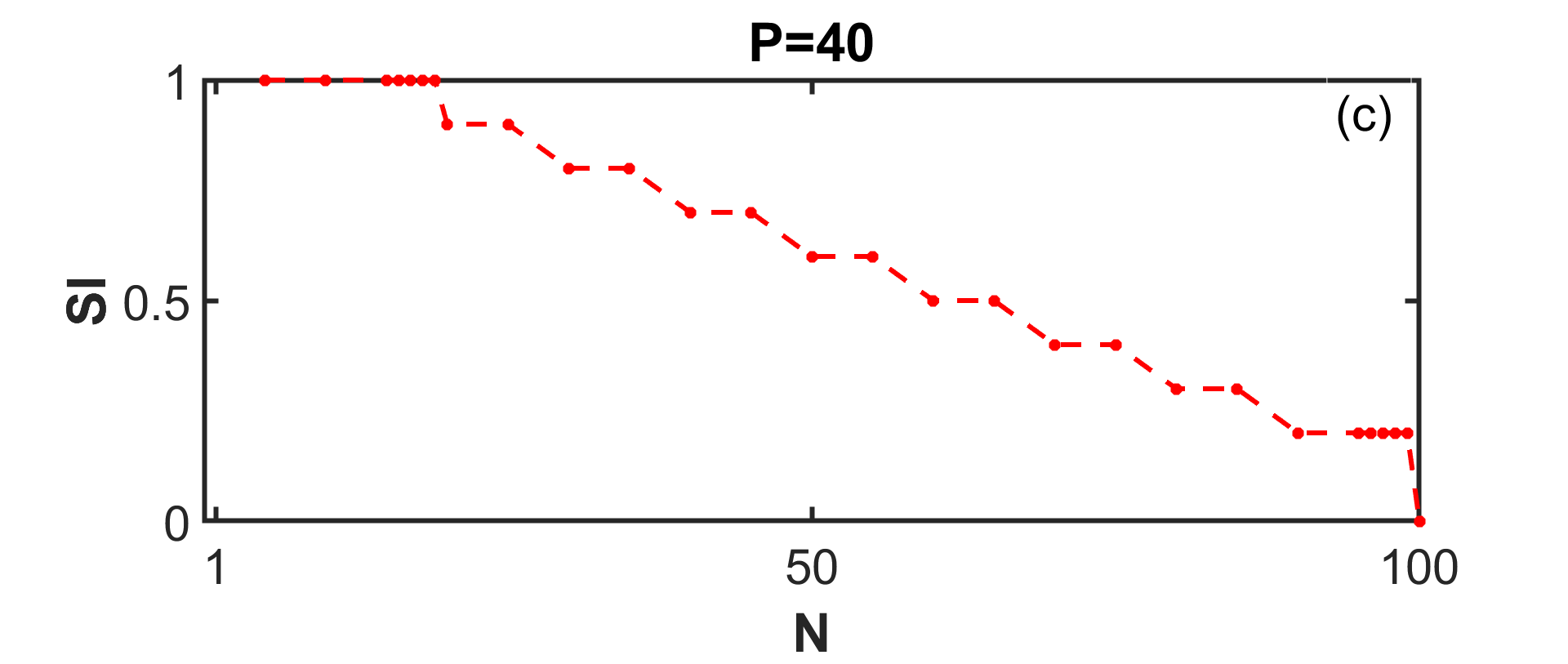}
	\includegraphics[width=7.5cm]{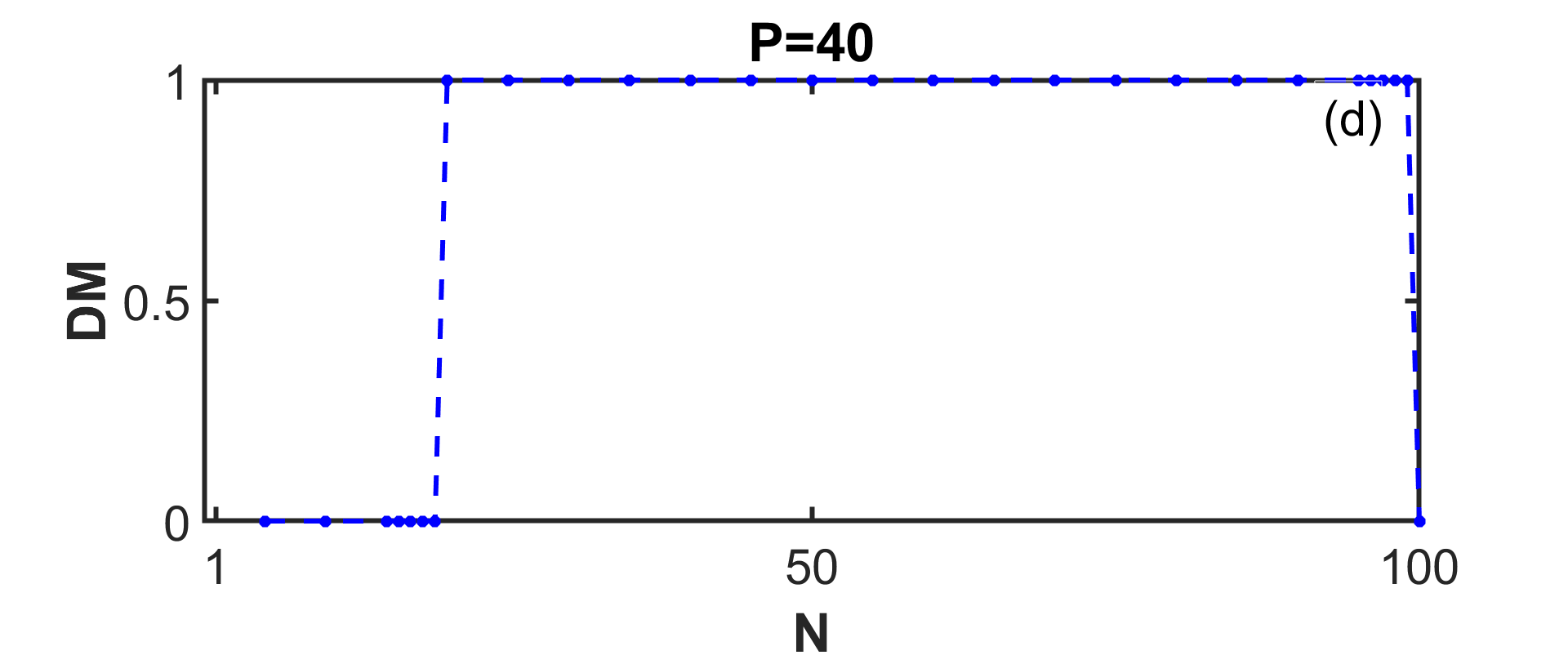}
	
	\caption{\label{fig.SGR6}
		Influence of the number of neurons $N$ subjected to the external electric field with the frequency $f=12$. (a) For $N=75$, spatiotemporal evolution for all the nodes, where we observe a chimera state for the first non-immersed $30$ neurons;
		(b) Imperfect traveling chimera in the non-submerged zone; (c) Strength of incoherence $SI$ and (d) Discontinuity Measure $DM$ as a function of $N$. Chimera state appears from $ N=20 $.} 
\end{figure}

\begin{figure}[!h]		
	\includegraphics[width=7.5cm]{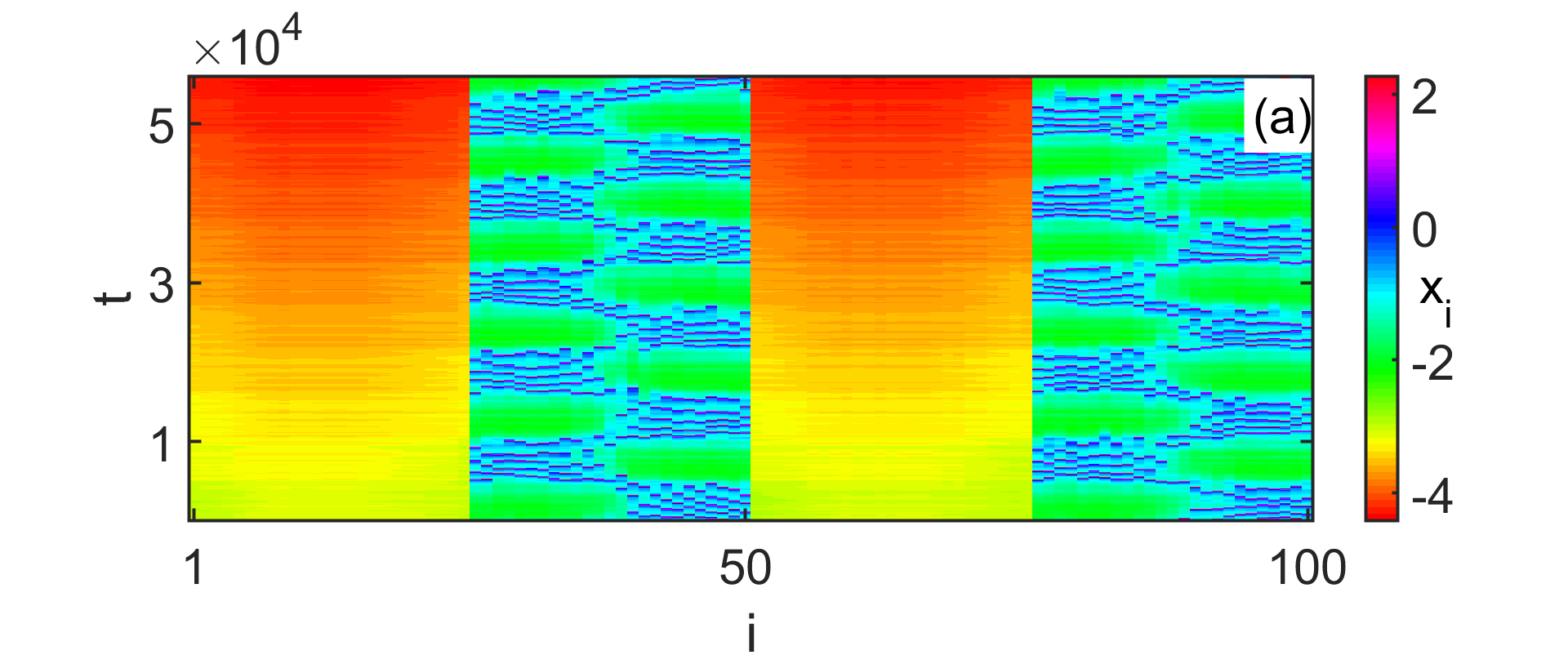}
	\includegraphics[width=7.5cm]{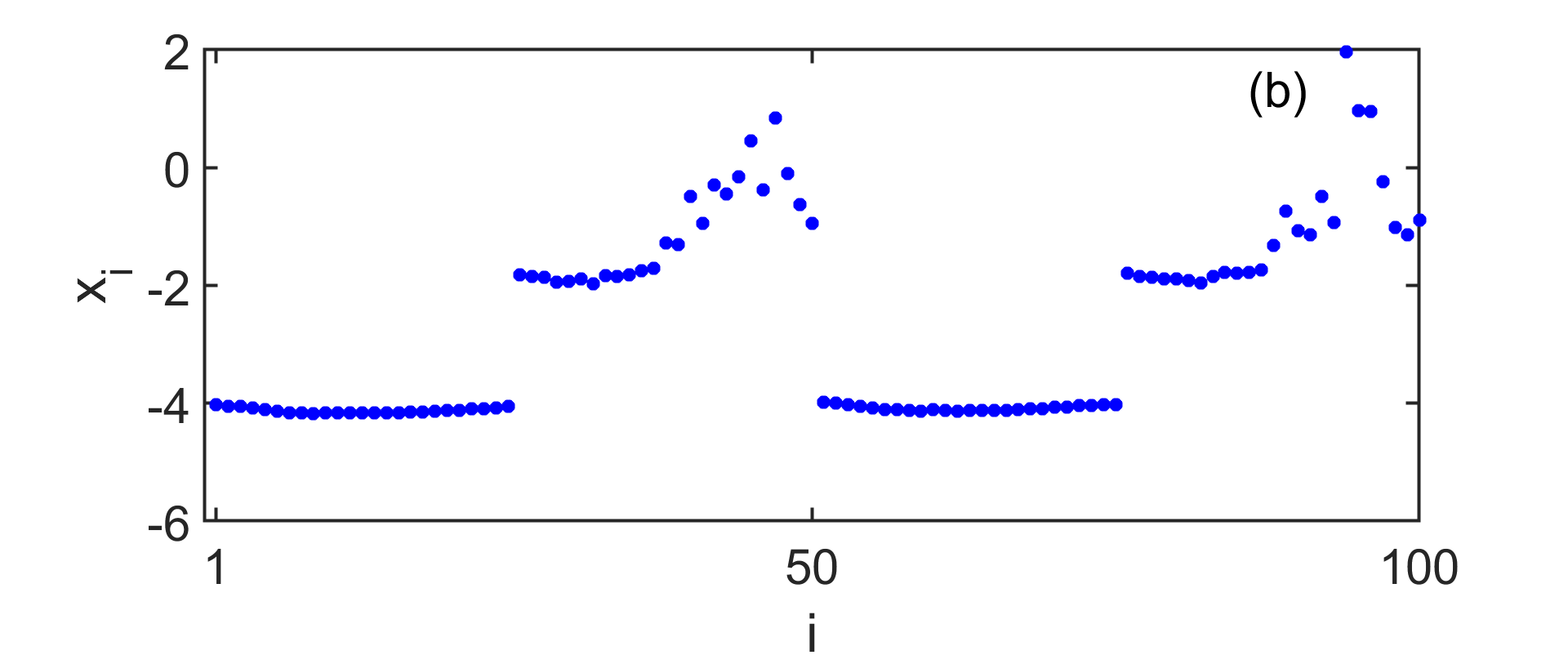}
	\caption{\label{fig.SGR7} (a) Multichimera state induced by application of an electric field in two non-consecutive regions; (b) Correspondent snapshot of the variables $x_i$.} 
\end{figure}

Now we apply the field on a large number of neurons decreasing in size, counted from the last elements of the ring.
We observe that as we vary $N$, the number of neurons subject to the electric field, different patterns emerge on the region nonsubject to the electric field. Figures \ref{fig.SGR6}(a) and (b) show the chimera states for $N = 75$ and $N = 30$. In the first case (\ref{fig.SGR6}(a)), the part of the network which is not under the action of the electric field appears to be incoherent. However, for $N = 30$ depicted in \ref{fig.SGR6}(b), we note a traveling chimera different from the state of \ref{fig.SGR1}(a), where the traveling structure is irregular.
A similar effect was reported by \cite{bera2016imperfect} and called imperfect traveling chimera. 
In order to study the influence of the number of neurons $N$ subjected to the external electric field, we calculated both $SI$ and $DM$ measures as a function of $N$. From Fig.\ref{fig.SGR6}(c) and (d), one can see that as the portion of neurons under the effect of the electric field increases, the value of $SI$ decreases until it reaches $0$, at this point $DM = 0$, confirming that the system is coherent when $N = M$. 
From the $DM$ and $SM$ measures, we verify that if the number of neurons under the influence of the electric field is below $19$ ($N\le19$), the entire network is in a state of total incoherence because $SI=1$ and $DM= 0$; the chimera state persists when $N$ takes values greater than $19$ and smaller than $99$ ($19 <N < 99$). For $N$ equal to $100$, the network enters a state of total coherence. It is worth mentioning that time-dependent chimera states, like the ones seen in Fig. \ref{fig.SGR1}(a) and 
Fig. \ref{fig.SGR6}(b), are read as incoherence by $SI$ and $DM$ measures. \\
  The simultaneous application of the electric field to unconnected parts of the same width of consecutive elements of the ring 
   leads the same to a manifestation of two coherent zones (those subjected to the field) intercalated by two other incoherent zones (Fig. \ref{fig.SGR7}(a)). 
We thus observe the formation of the multichimera state, as
confirmed by the $SI = 0.6$ and $DM = 2$ measures. 
The distribution of $x$ along the ring at an arbitrary time, depicted in Fig. \ref{fig.SGR7}(b), stresses these features. Moreover, we note that there is also a division between coherent and incoherent domains inside the so-called incoherent parts, similar to the alternating chimera state of Fig. \ref{fig.SGR5}(a). It appears that the application of an electric field gives place to the nesting of chimeras of different types, a phenomenon that should be studied in detail, but we leave it here since this is not within the scope of this work.
 
\begin{figure}[h!]
	\begin{minipage}[c]{.46\linewidth}
		\includegraphics[height=2.6cm,width=4.5cm]{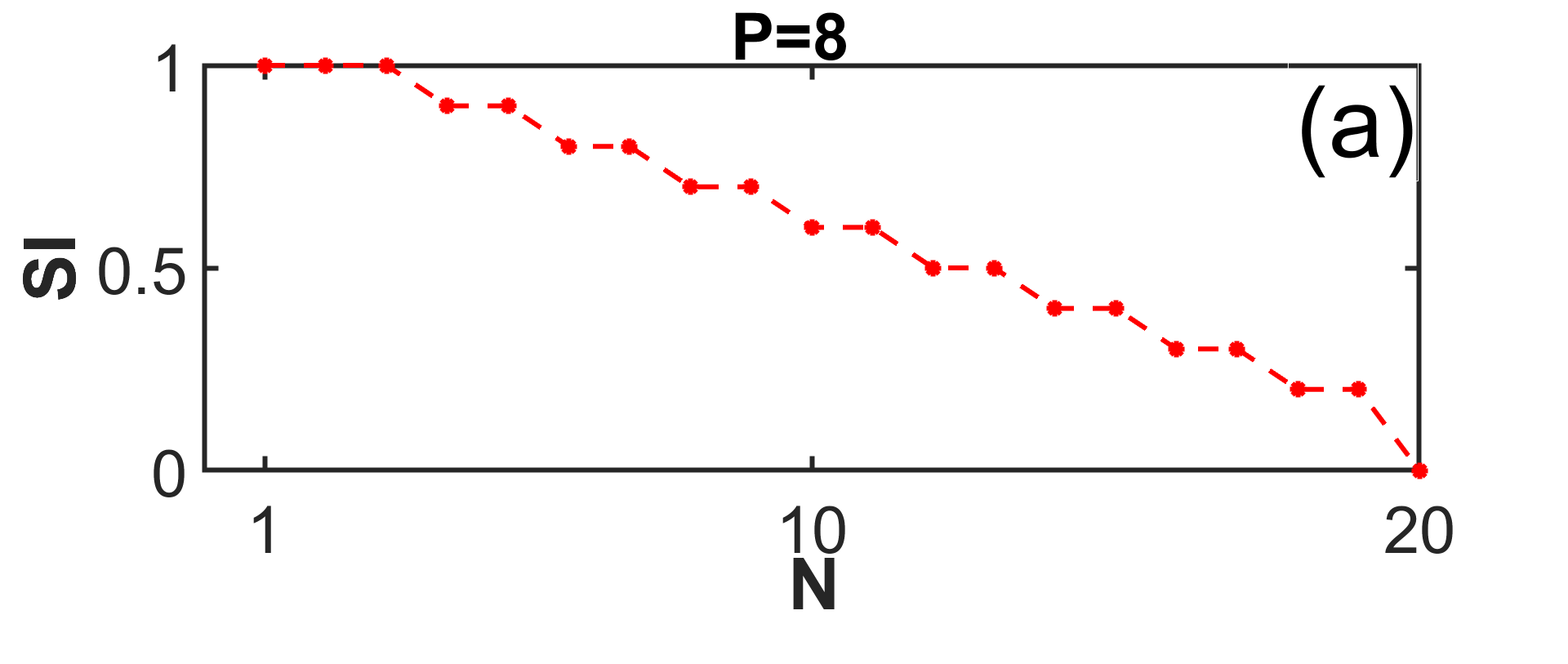}
		\includegraphics[height=2.6cm,width=4.5cm]{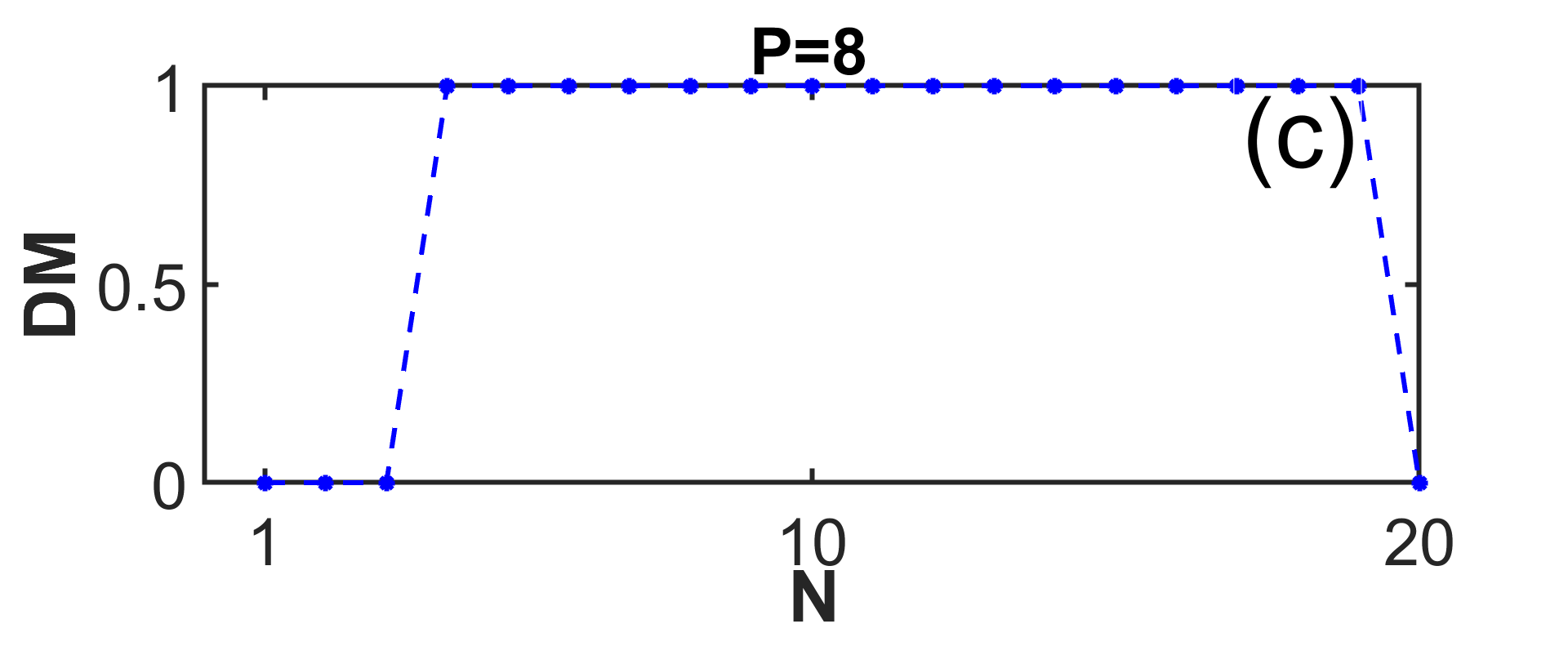}
		\includegraphics[height=2.6cm,width=4.5cm]{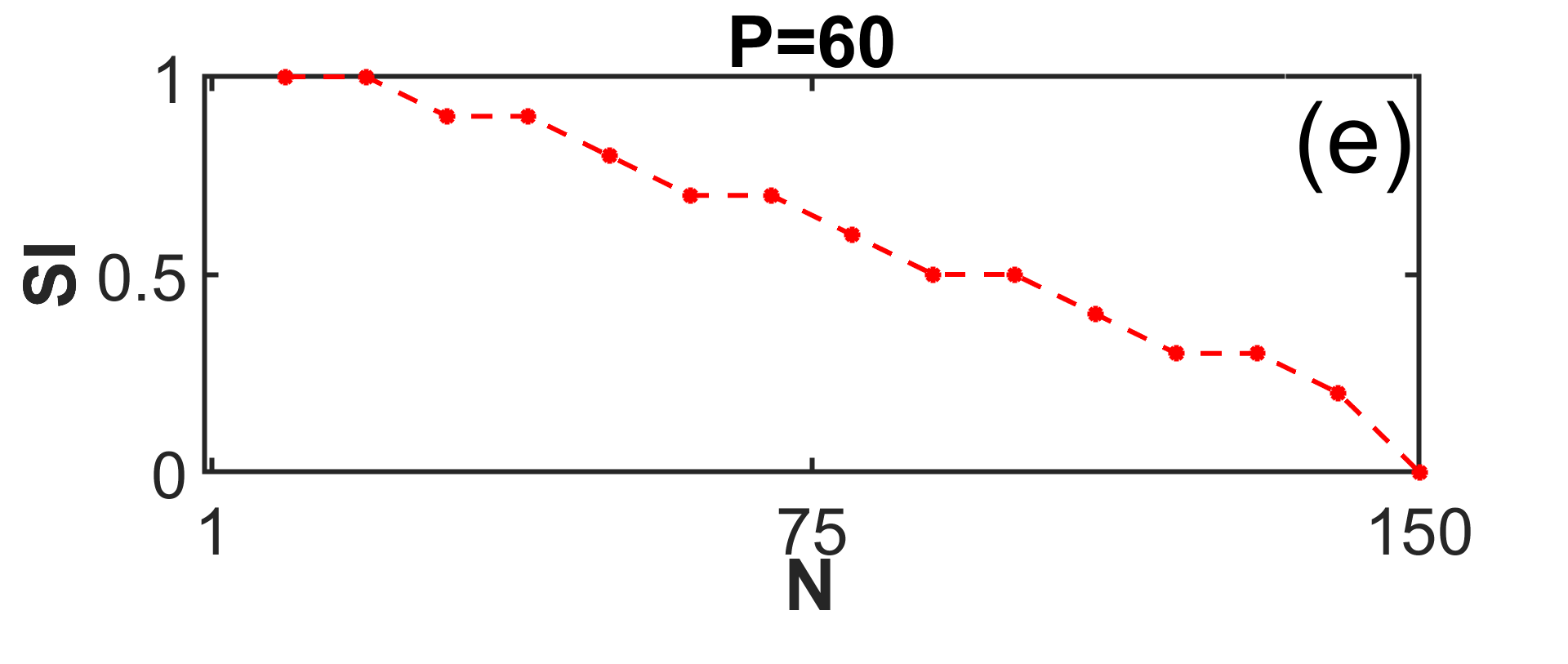}
		\includegraphics[height=2.6cm,width=4.5cm]{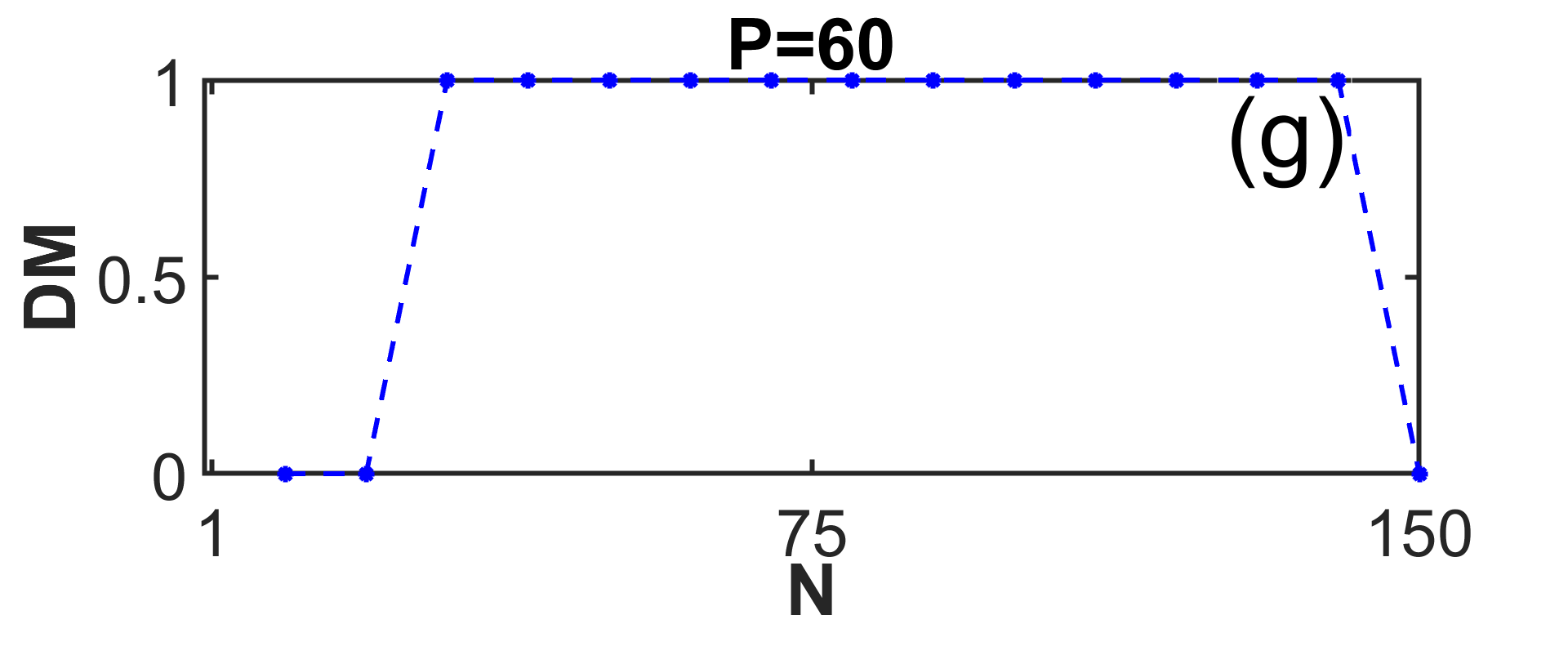}
	\end{minipage} \hfill
	\begin{minipage}[c]{.46\linewidth}
		\includegraphics[height=2.6cm,width=4.5cm]{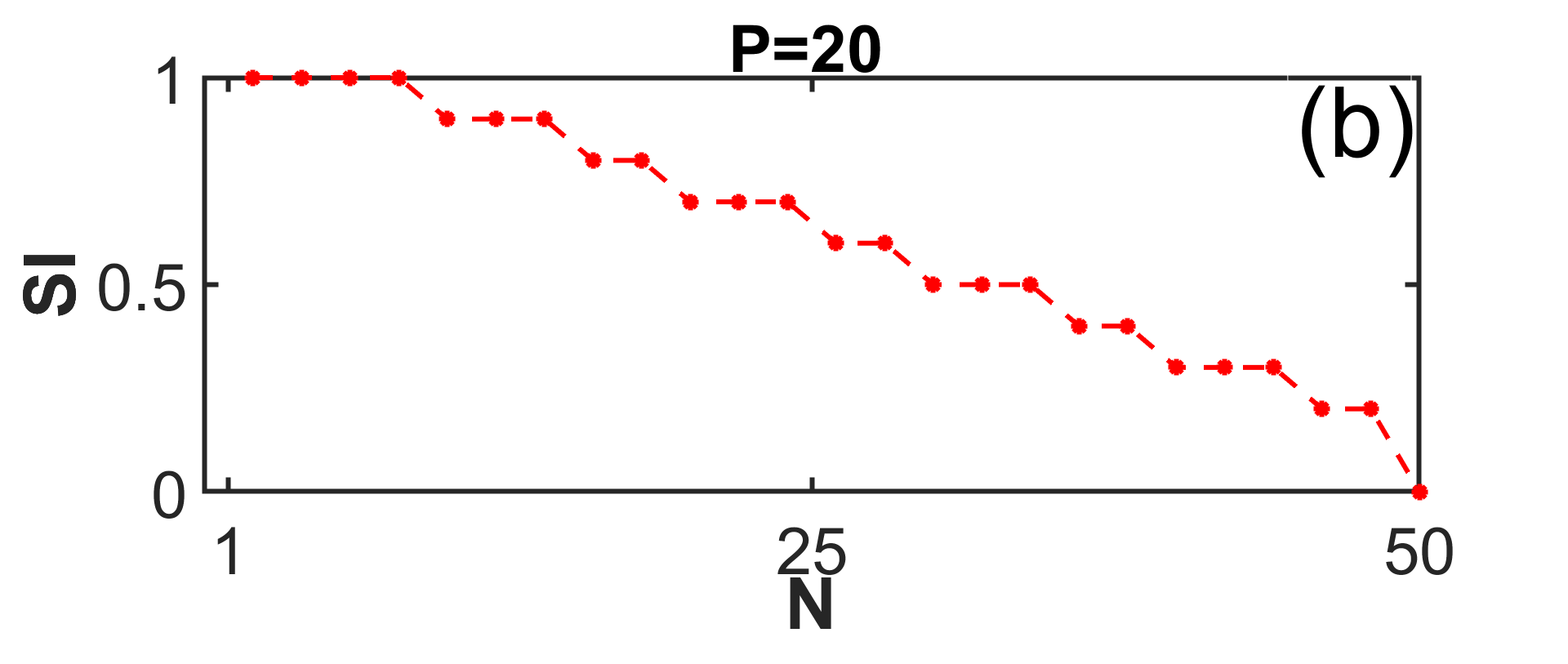}
		\includegraphics[height=2.6cm,width=4.5cm]{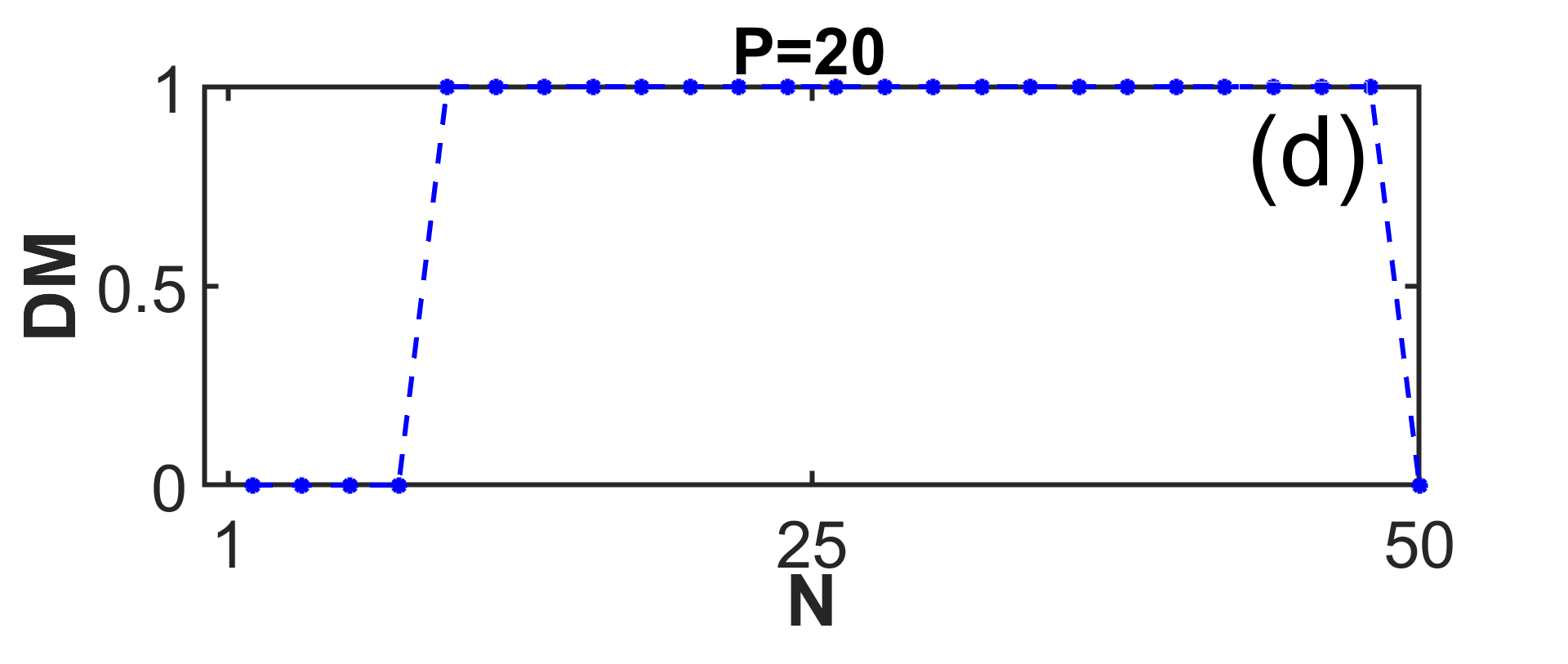}
		\includegraphics[height=2.6cm,width=4.5cm]{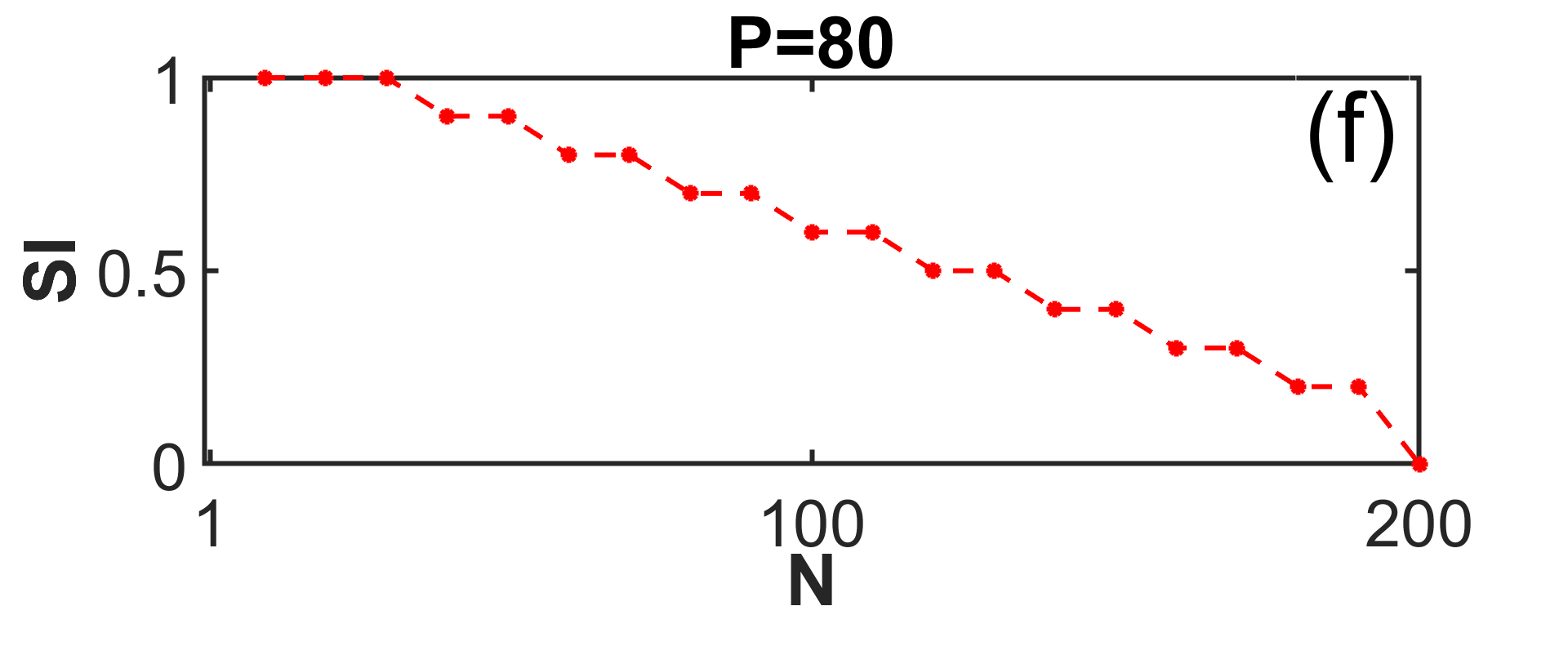}
		\includegraphics[height=2.6cm,width=4.5cm]{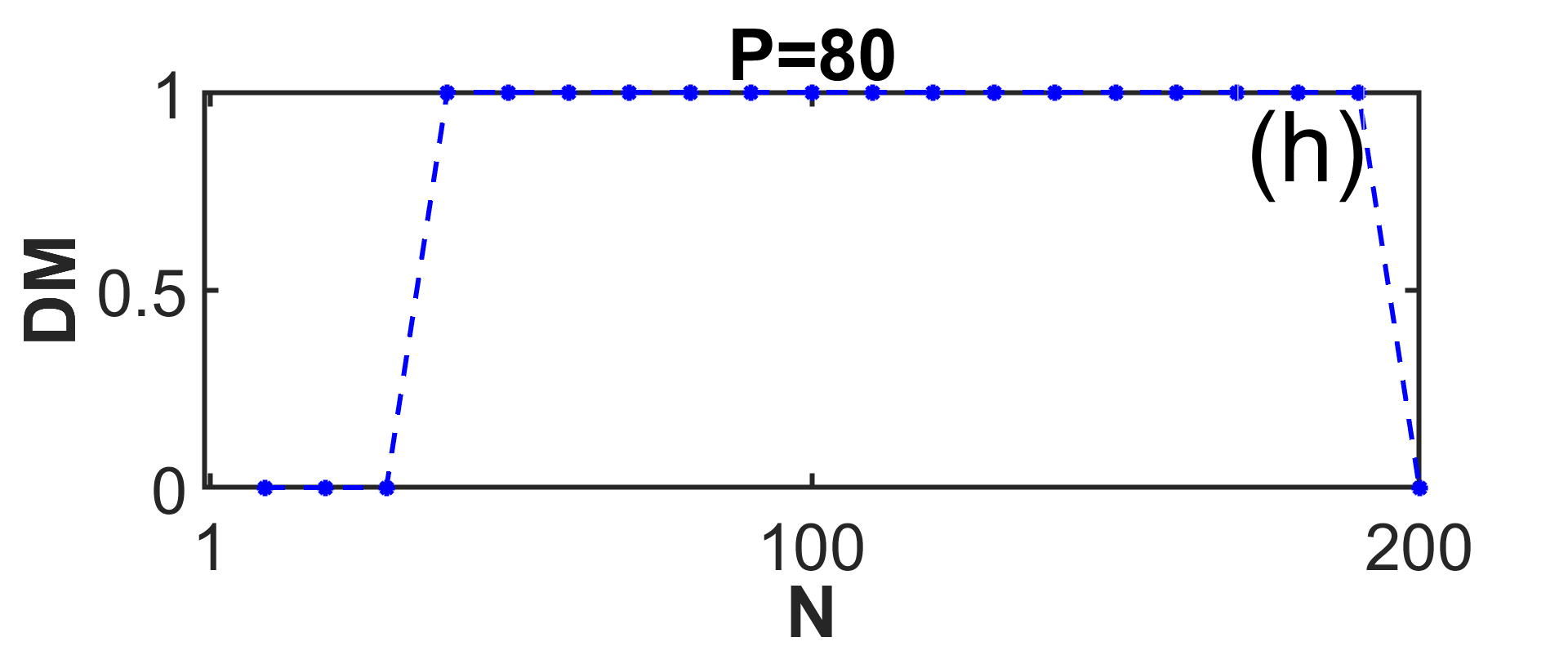}
	\end{minipage}
	
	\caption{\label{fig.SGR8} 
		Influence of the number of neurons $M$ of the whole ring for $f=12$. Strength of incoherence and discontinuity measure for (a) \& (c) $M = 20$, (b) \& (d)$M = 50$, (e) \& (g)$M = 150$, (f) \& (h)$M = 200$. Appearance of the chimera state from 20\% of the total number of neurons in the ring.}
\end{figure}

To verify the generality of our results, we carried out a study for other values of $M$. The analysis of these different cases through the $SI$ and $DM$ measures (Fig. \ref{fig.SGR8}) reveals that the chimera state appears in the network from a certain fixed proportion of the elements embedded in the electric field. 
Indeed, for $M$ values, respectively equal to $20$, $50$, $150$, and $200$, the chimera state appears when the number of neurons $N$ immersed in the electric field reaches the values $4$, $10$, $30$, and $40$ respectively. 
That is to say, a proportion of at least 20\% of the total number $M$ of neurons in the ring, immersed in the electric field, are capable of inducing the chimera state. However, when the field drowns out the whole set, the network elements enter a state of total coherence. To calculate SI and DM we used $\delta = 0.02*|(x_{max}-x_{min})|$ \cite{dogo,gopal2014observation}.

\begin{figure}[h!]
	\includegraphics[width=7.5cm]{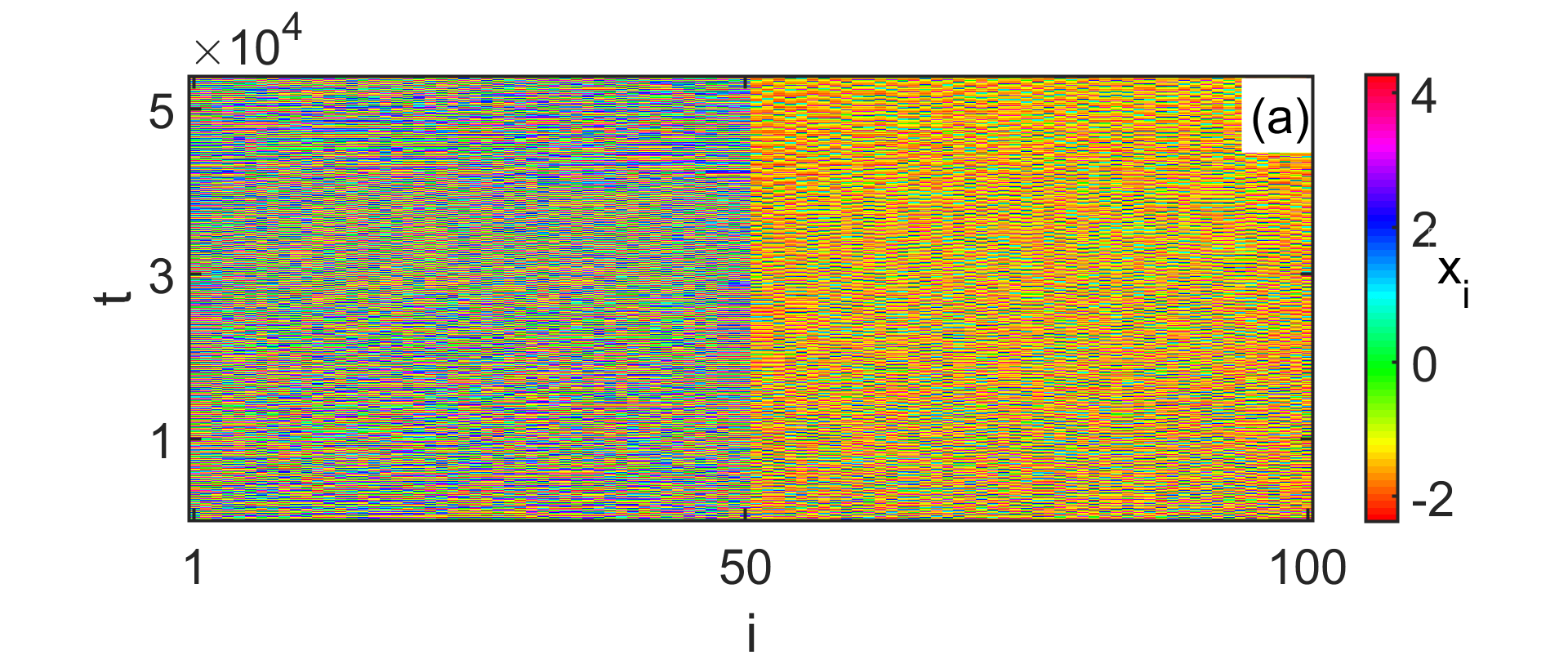}	
	\includegraphics[width=7.5cm]{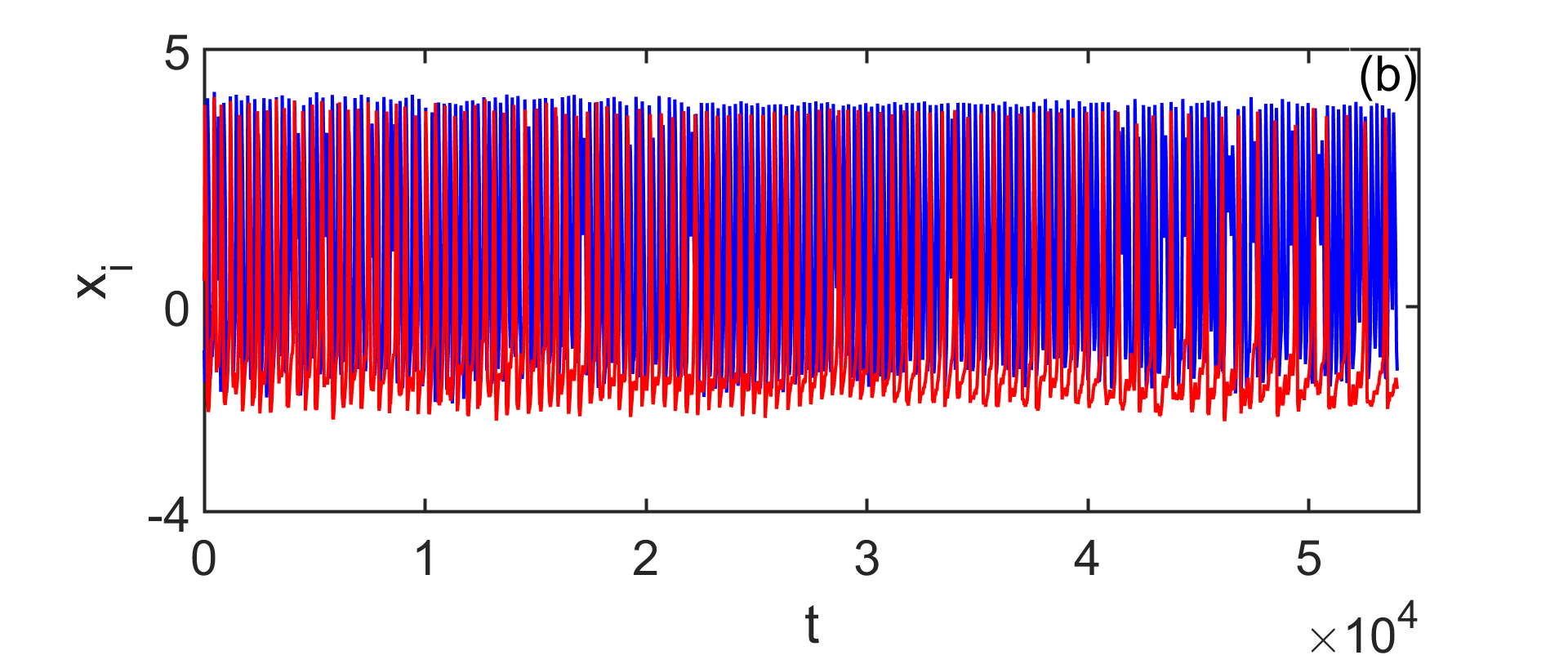}	
	\caption{\label{fig.SGR9} 
		Incoherent state induced by $I=35$. (a) Spatiotemporal evolution of the $x$-variables; (b) Time series showing irregular bursting. The red curve is for the neurons of the part immersed in the field, and blue for the part not immersed in the electric field.}
\end{figure}

We checked the results shown in Figs \ref{fig.SGR6}, \ref{fig.SGR7} for several values of the intensity and frequency of the applied field and did not find any variations for SI and DM. \\
The previous results are obtained only for small values of the external current; in this case, $I = 3.5$. We modify the latter by assigning it the value $35$. It appears that, regardless of the area sprayed by the electric field, all the elements of the network remain in an incoherent, regular bursting state, shown in Fig. \ref{fig.SGR9}(a) and supported by Fig. \ref{fig.SGR9}(b) with the temporal evolution of the elements in the two domains, where we note that the red curve of the area influenced by the electric field evolves into a state of sparse burstings and a progressive decrease of their amplitude. Thus we witness the state's destruction shown in Fig. \ref{fig.SGR4} when we increased the chemical coupling strength to ten $(k_4 = 10)$.

\begin{figure}[h!]	
	\includegraphics[width=7.5cm]{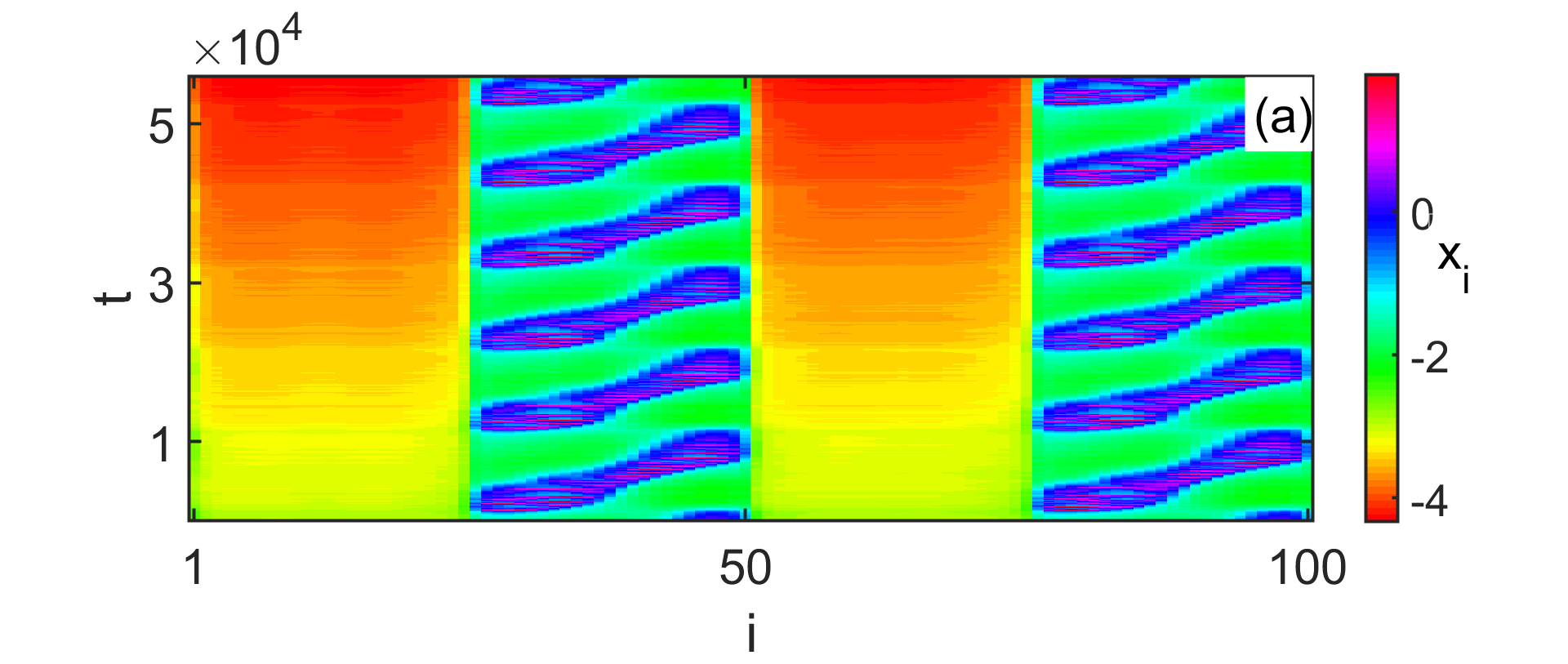}	
	\includegraphics[width=7.5cm]{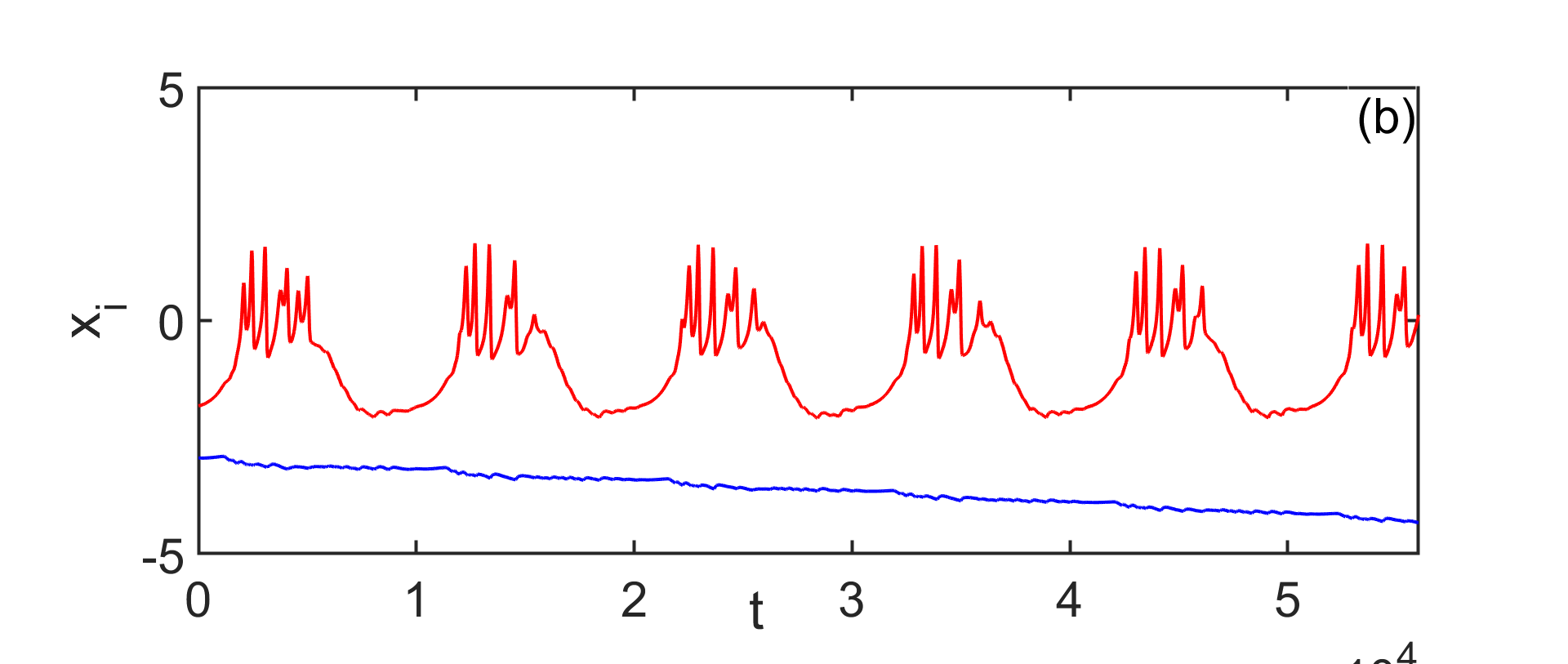}	
	\caption{\label{fig.SGR10} 
		Impact of electrical coupling ($k_3=2$ and $k_4=3$) for $f=12$. (a) Spatiotemporal evolution of 
		the $x$-variables shows a \textit{multicluster traveling chimera}. (b) Time series shows chaotic bursting.
		The blue curve is for the neurons of the part immersed in the field and red for the part not immersed in the electric field.}
	
\end{figure}

All of the above was done in the absence of electrical coupling. To take this into account, we decide to use the values for which a traveling chimera appeared. Therefore we apply the field in two different regions for the coupling strengths $k_3 = 2$ and $k_4 = 3$. We observe the appearance of two blocks where there is a manifestation of traveling chimera(Fig.\ref{fig.SGR10}(a)). These blocks correspond to the zones not subjected to the electric field. The areas in which the traveling chimera takes place are separated by those subject to the electric field. Additionally, the simultaneous presence of these two blocks, the seat of traveling chimera state, has led us to qualify this phenomenon as \textit{ multicluster traveling chimera}. These areas subject to the electric field have not changed dynamics. Then we can say that for this specific case, the introduction of an external electric field has the ability to split a ring exhibiting a traveling chimera state into two subsets exhibiting this same phenomenon, where each element in the traveling domain remains in a chaotic bursting regime \cite{kaas,hansel}. (Fig.\ref{fig.SGR10}(b)).This state of chaos is confirmed by the value of the correlation coefficient produced by the 0-1 test, which is 0.9906 \cite{Georg A}. For the 0-1 test to be complete, in addition to the correlation coefficient, which must be close to 1, the translation variables must describe a Brownian motion (or unbounded), and the mean square displacement  must evolve asymptotically along a straight line, which is shown in Fig.\ref{fig.SGR13} of Appendix\ref{Ap2}. 

\section{Conclusion}

We study a neuronal network composed of HR neurons linked together by local electrical couplings and non-local chemical couplings. Considering the electric field of each neuron, we verify the influence of an external electric field on the chimera states that the network can exhibit. First, taking no electrical coupling, we verified the presence of traveling chimera states, as presented in the work of Mishra \textit{et al.} \cite{mishra2017traveling}.  
In addition to that, by increasing the value of the external current to $I= 35$, we found a novel phenomenon that we called \textit{traveling multicluster chimera breather}, which is marked by the presence of multiple nonstationary clusters of coherence. Moreover, by strengthening the chemical synaptic coupling, we obtained a \textit{multicluster chimera breather}, a result similar to the one of Zhu et al. \cite{zhu2012}.  We used the local order parameter to measure the coherence of such states.

Next, to introduce an external electric field on the network, we followed the model proposed by Ma \textit{et al.} \cite{ma2019model} and adapted the HR model. Then we introduce an external electric field of the form $E_{ext}=E_m\sin(ft)$ on fixed portions of the ring and vary the electric field's frequency. For small values of $f$ ($f=0.01$), we found that the external electric field destroys the previous traveling chimera states, 
giving place to an alternating chimera state (see Fig.\ref{fig.SGR5}(b)). Moreover, the spikes' amplitude and frequency inside the burst appear larger for the neurons interacting with the external electric field $E_{ext}$. 
Following this, increasing the frequency to $f=12$ leads us to a state where the neurons' activity under the influence of $E_{ext}$ seems to be totally suppressed. The portion of the network not submitted to $E_{ext}$ exhibits  an imperfect traveling chimera state (Fig.\ref{fig.SGR6}(b)).

To distinguish between different types of chimera states, we choose to use the strength of incoherence ($SI$) and discontinuity measures ($DM$). Then we varied the network size, the number of neurons subjected to $E_{ext}$ and verified the effect of having two portions of the network subjected to $E_{ext}$. The network then presents chimeras, imperfect traveling chimeras, and multichimeras, depending on how $E_{ext}$ is applied. More precisely, we showed that we need a minimum of $20\%$ of the $M$ neurons in the network to be under the electric field's influence to exhibit a chimera state. Not less important, for a higher value of external current (I = 35), we showed that the network is split into two incoherent states, with irregular bursting.

 Finally, we included the electrical coupling and verified the presence of \textit{multitraveling chimera states} when the network is submitted to $E_{ext}$ in two separated regions. And again, the neurons subjected to $E_{ext}$ seemed to be in a state of progressive supression.

\section*{Acknowledgements}
	
HAC thanks ICTP-SAIFR and FAPESP grant 2016/01343-7 for partial support. RPA acknowledges financing  from Coordena\c{c}\~{a}o de Aperfei\c{c}oamento de
Pessoal de N\'{i}vel Superior - Brasil (CAPES), Finance Code 001.

\section*{Appendix: Influence of $p$: characterization using the traveling speed measurement method And Application of the 0-1 test method for the determination of chaos.}

\appendix\subsection{Influence of $p$: characterization using the traveling speed measurement method.}\label{Ap1}
Regarding the role of $p$, first, we consider the absence of electrical coupling ($k_3 = 0$,  $k_4=9$, $I=3.5$) and let $p$ vary between 2 and 49  since $p$ cannot be smaller than 2 to avoid local coupling nor larger than $ (M / 2)-1$  not to repeat the chemical connections. For all these values of $p$, we always witness the manifestation of traveling chimera (see Fig.\ref{fig.SGR11}.(a,b,c)). Then we set the electrical coupling coefficient to $1$ ($k_3 = 1$) and keep the chemical coupling coefficient at $9$ $(k_4 = 9)$; then vary $p$ from $2$ to $49$. This approach has shown that the ring keeps the traveling chimera in only one direction (patterns oriented from right to left), except for $p = 15$, where the direction of propagation changes (patterns oriented from left to right), as shown in Fig.\ref{fig.SGR11}.(d,e,f).\\
 Hizanidis et al. \cite{Hizanidis1} proposed a technique for identifying the traveling chimera and its traveling velocity.  Given that the incoherence zone is not fixed and moves along the ring, one way to distinguish the traveling chimera is to locate the node of the ring which carries the maximum amplitude $x_{max}(t)$ or minimum $x_{min}(t)$ at each instant. Thus, the indices of the nodes of maximum or minimum amplitude are recorded at each moment in a vector $J_{max}(t)$ such that for our time series, we have:
\begin{figure}[h!]
	\begin{minipage}[c]{.46\linewidth}
		\includegraphics[height=2.6cm,width=4.5cm]{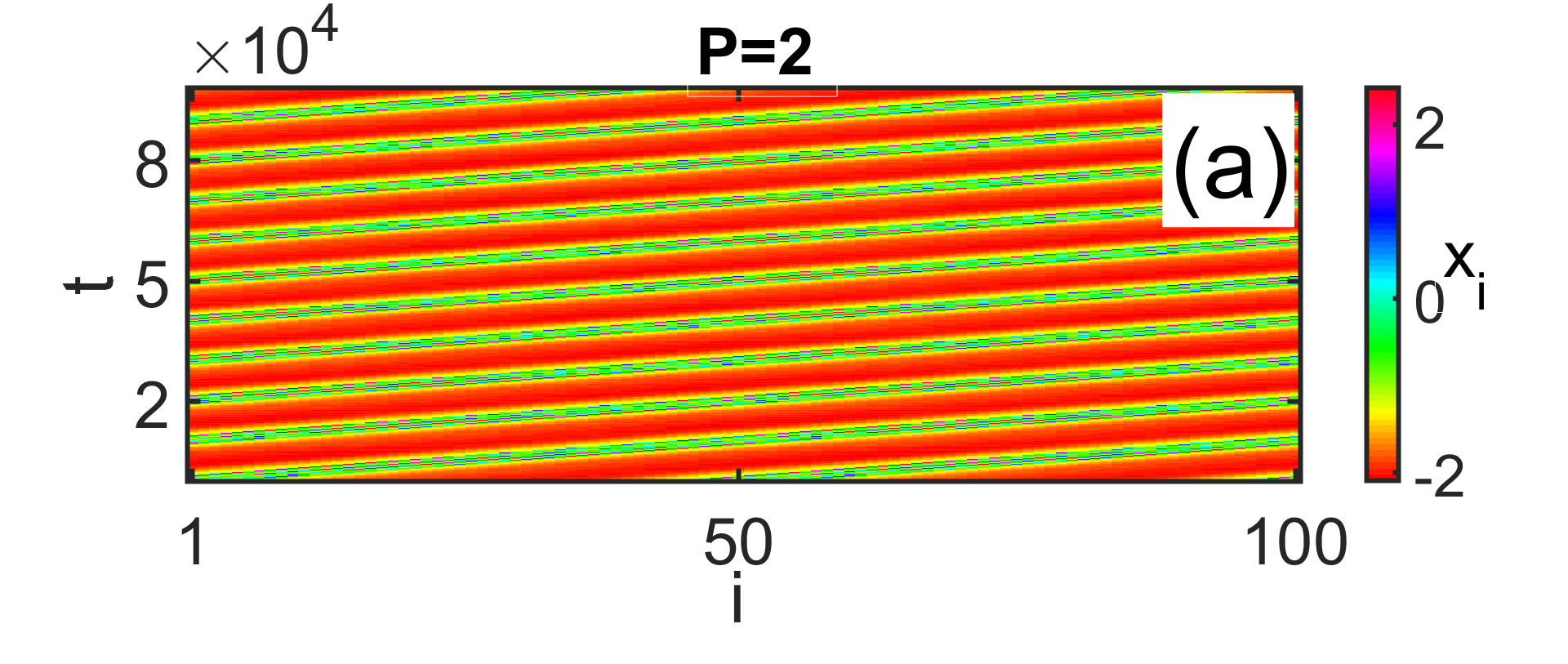}
		\includegraphics[height=2.6cm,width=4.5cm]{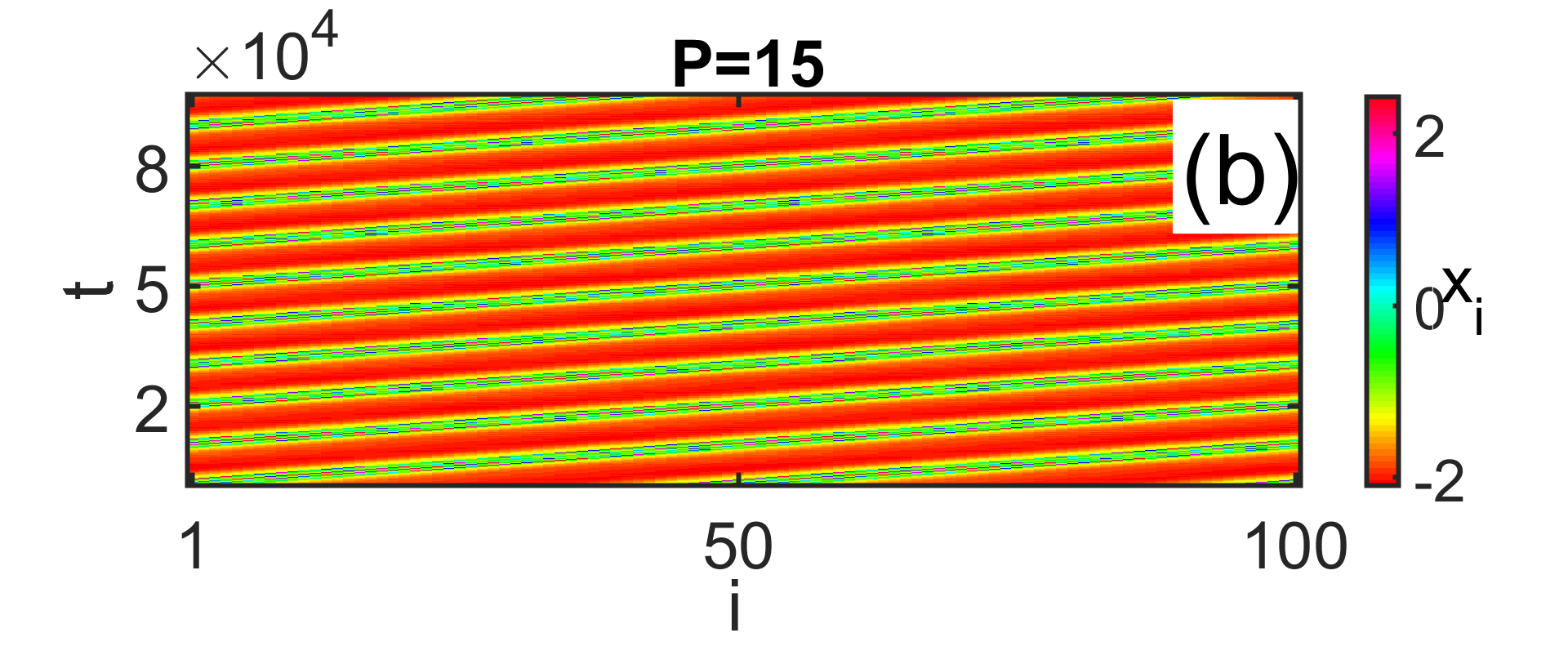}
		\includegraphics[height=2.6cm,width=4.5cm]{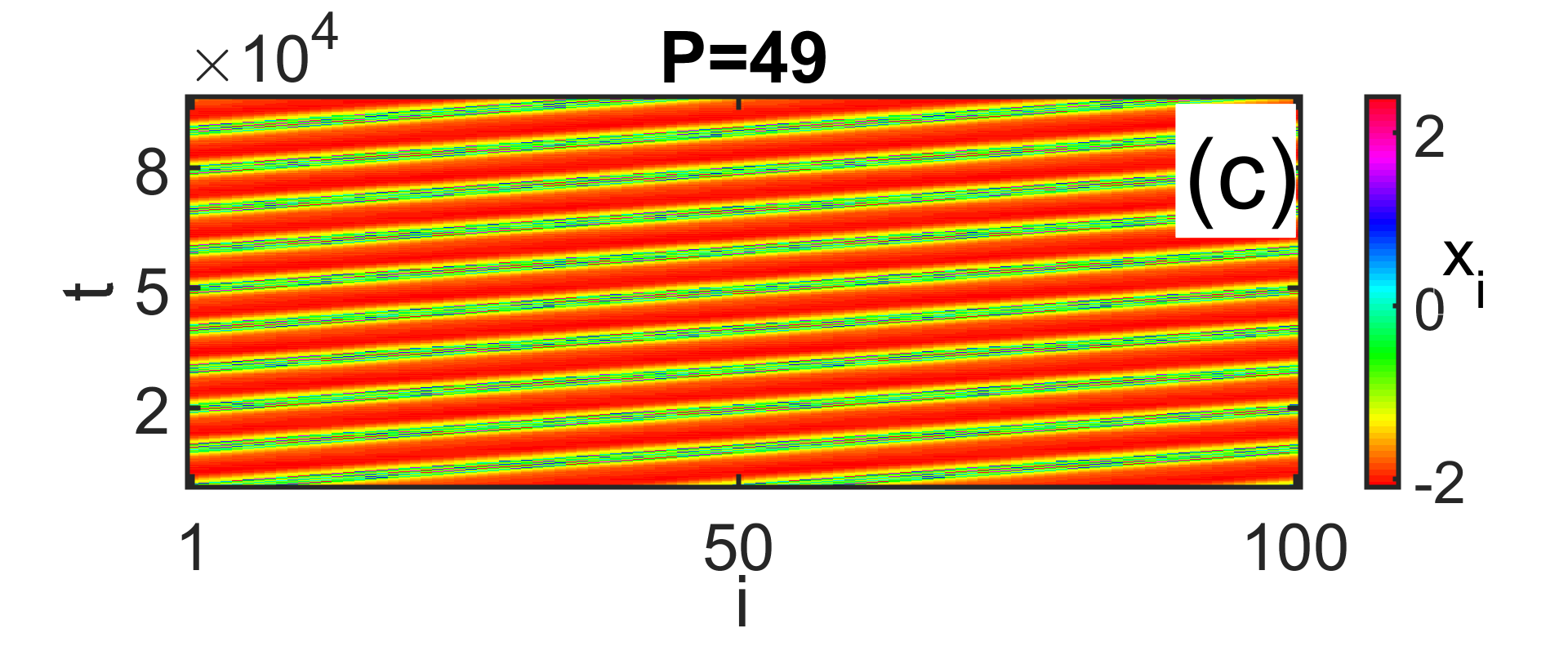}

	\end{minipage} \hfill
	\begin{minipage}[c]{.46\linewidth}
		\includegraphics[height=2.6cm,width=4.5cm]{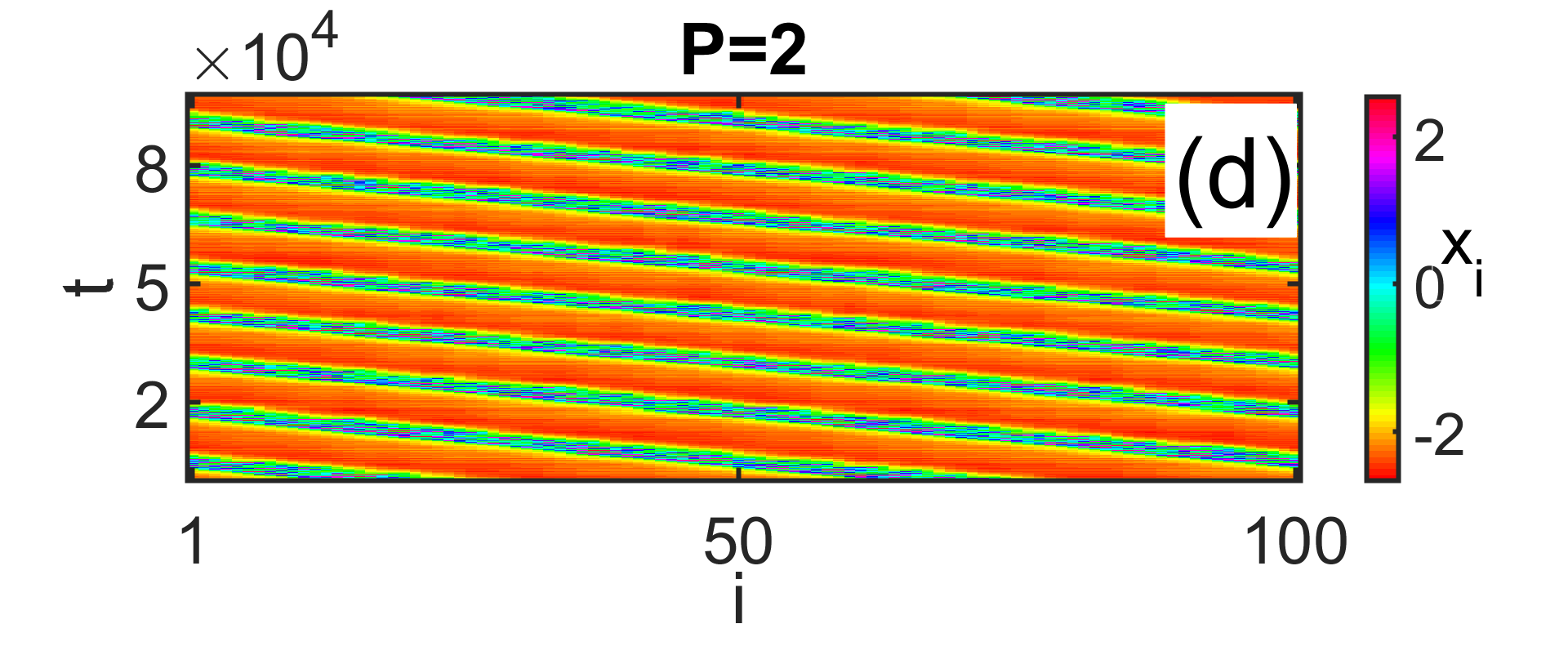}
		\includegraphics[height=2.6cm,width=4.5cm]{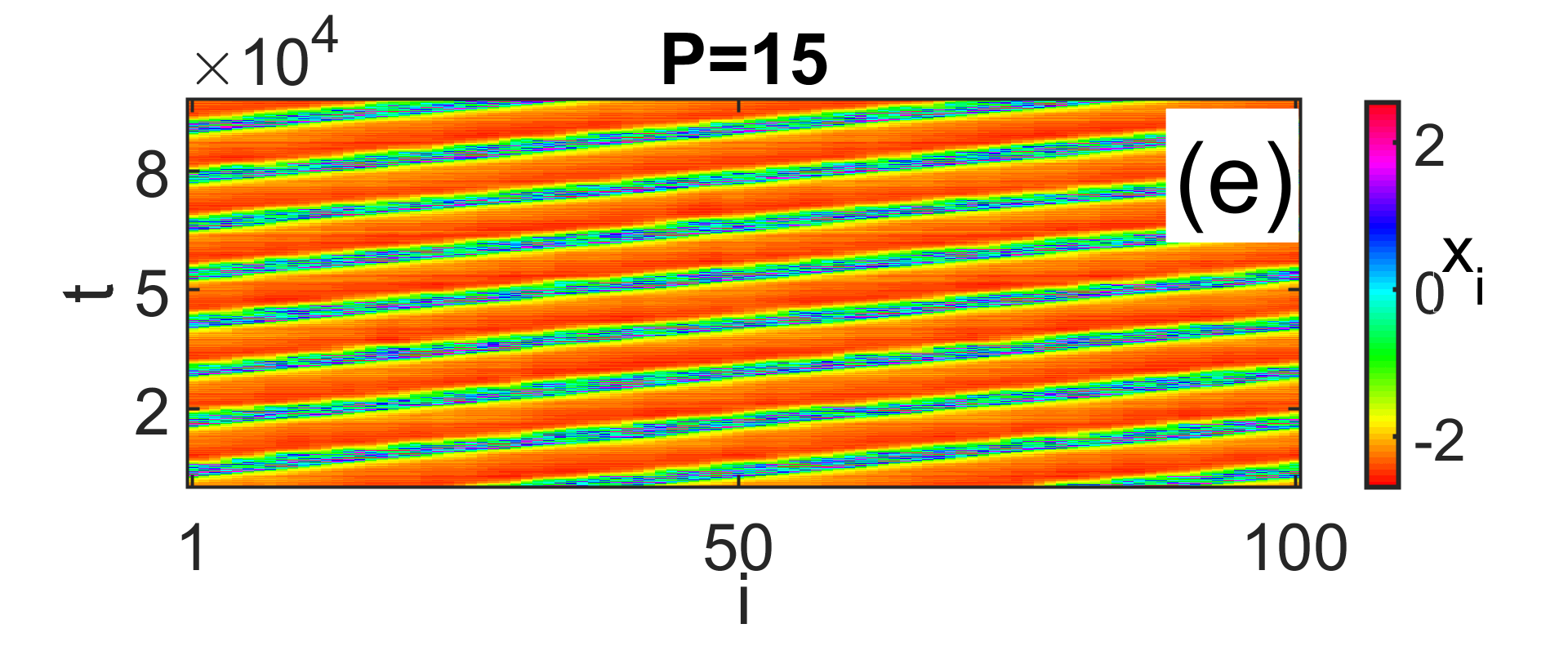}
		\includegraphics[height=2.6cm,width=4.5cm]{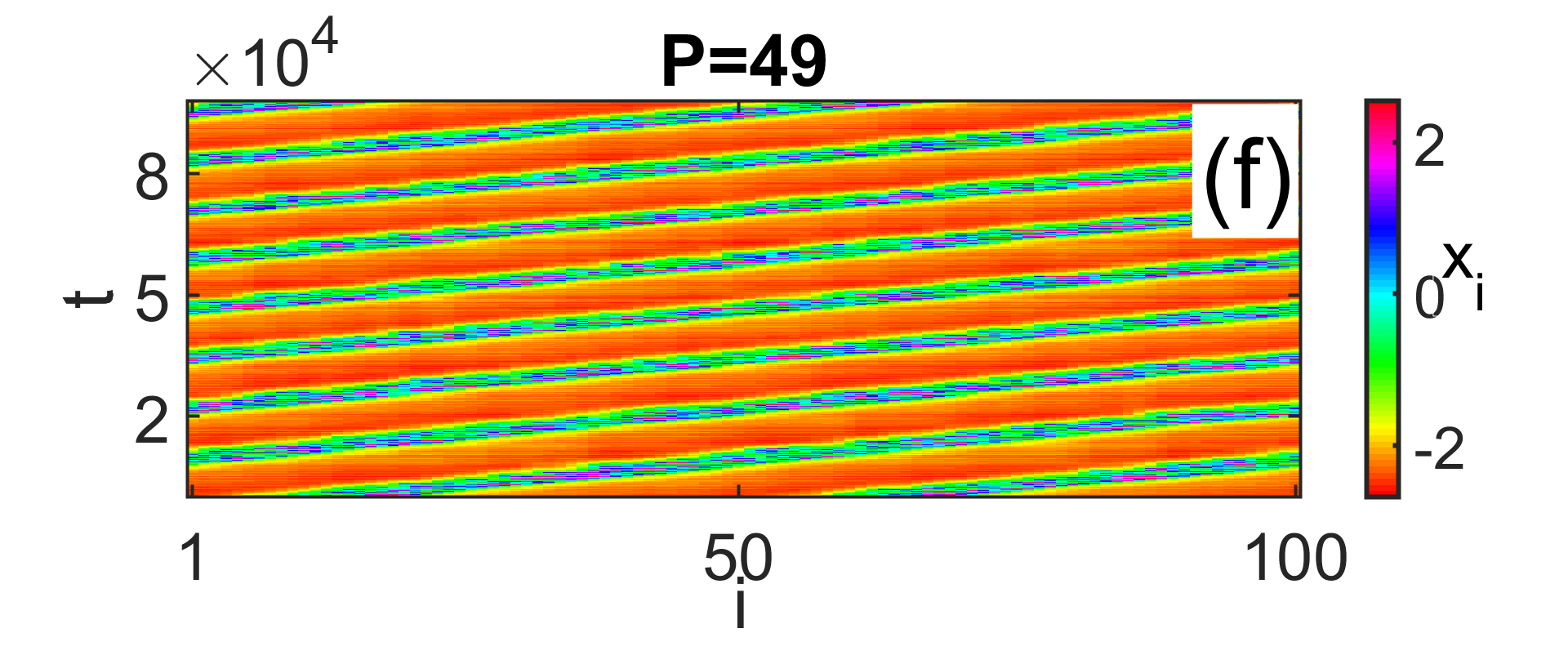}
	\end{minipage}
	
	\caption{\label{fig.SGR11} 
		Role of the number of neighbors $p$ on the dynamics of the network's traveling chimera (a), (b), (c) without electrical coupling $(k_3=0, k_4=9)$; (d), (e), (f) with electrical coupling $(k_3 = 1, k_4 = 9)$.}
\end{figure}

\begin{equation}
	J_{max}(t)=\max\{x_1(t),x_2(t),..., x_N(t)\}.
\end{equation}
 Analogously, the instantaneous indices of the minimum amplitudes $J_{min}(t)$ can be recorded.
Because of the displacement of the chimera state, each element is supposed  to carry the maximum or minimum amplitude successively. Therefore, the temporal representation of $J_{max}(t)$ or $J_{min}(t)$ will let appear a certain periodicity of behavior. Thus, the period T necessary for any element to carry the maximum or minimum amplitude can be obtained by Fourier transform of $J_{max}(t)$ or $J_{min}(t)$. The largest period $T_{tr}$ corresponding to the smallest frequency $f_{tr}$ of the large amplitude of the frequency spectrum of $J_{max}(t)$, makes it possible to calculate the traveling speed of the $N$ elements of the ring in the form:
\begin{figure}[h!]	
	\includegraphics[width=7.5cm]{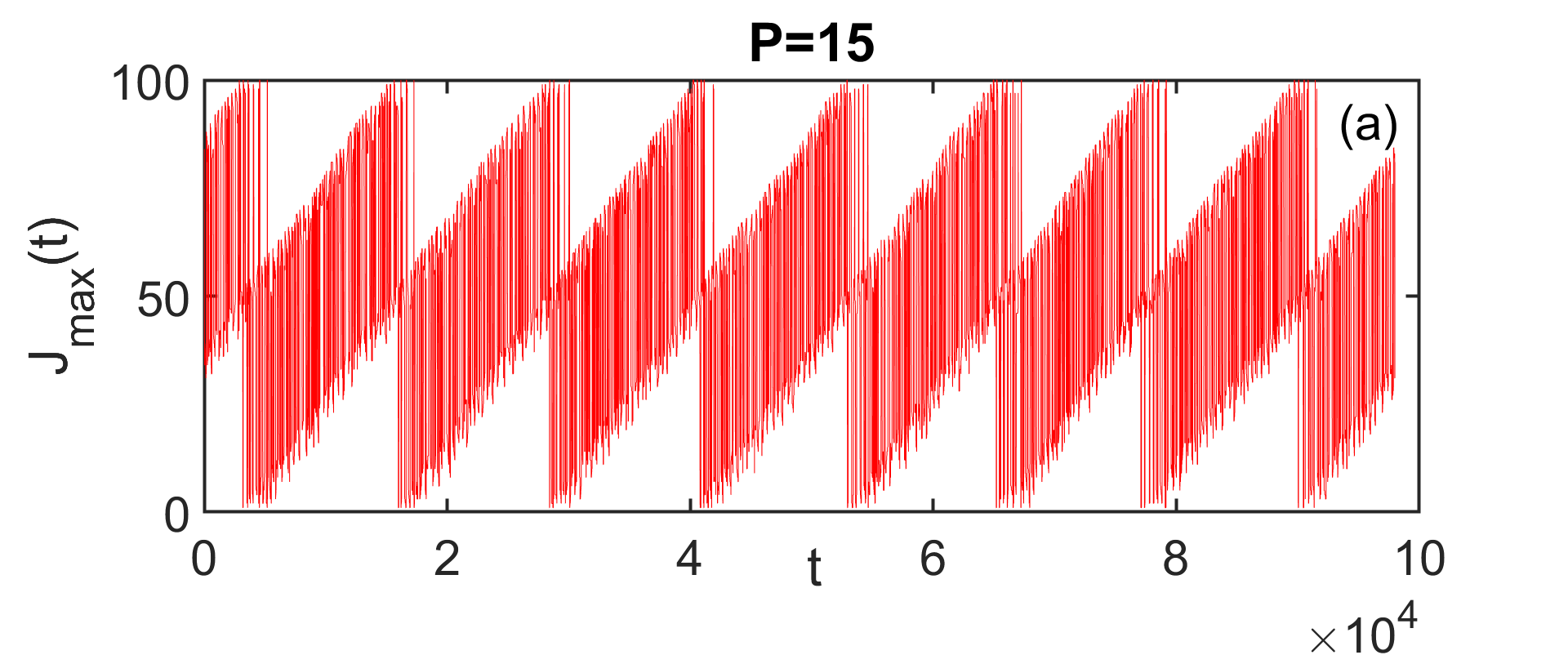}	
	\includegraphics[width=7.5cm]{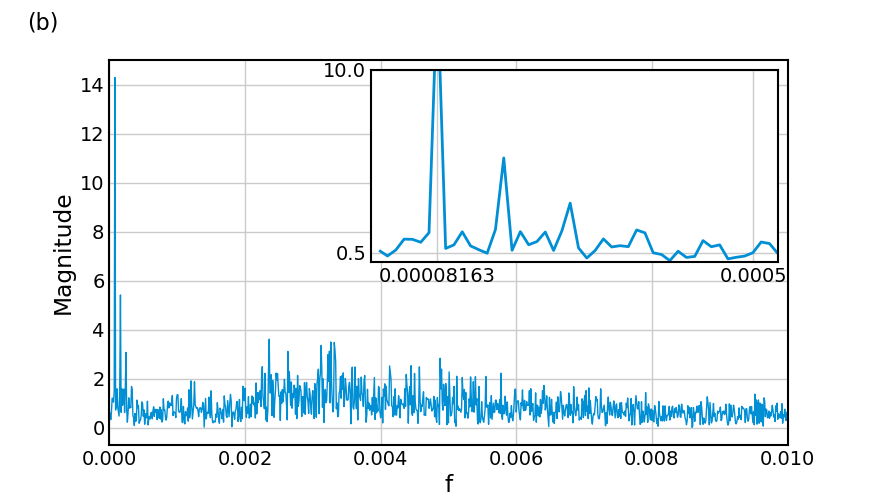}	
	\caption{\label{fig.SGR12} 
		a) Position of the oscillator displaying maximum $x$-value with time for $p=15$ for $k_3 = 1$, $k_4 = 9$.  The periodicity of the waveform refers to that of the traveling chimera state. (b) Fourier transforms of $J_{max}(t)$, with low frequencies, expanded in the inset, to discern the frequencies  of the traveling motion around the ring ($f_{tr}=0.0000816$).}

\end{figure}

\begin{equation}
	v_{tr}=N/T_{tr}=Nf_{tr}.
\end{equation}.

We continue by evaluating the propagation speeds of traveling in the ring for each of the values of $p$. It appears that in the absence of the electrical coupling ($k_3 = 0,$ $k_4 = 9$), the speed of the traveling is 0.0102 ($v_{tr} = 0.0102$) while in the presence of the electrical coupling ($k_3 = 1,$  $k_4 = 9$) the speed of the traveling becomes 0.00816 ($v_{tr} = 0.00816$). It implies that the introduction of electrical coupling can slow down the chimera traveling. An example of $J_{max}$ representation and its Fourier spectrum for $p = 15$ is represented in figure 12. We can observe that $J_{max}(t)$ is periodic (Fig.\ref{fig.SGR12}(a). And the analysis of the frequency spectrum (Fig.\ref{fig.SGR12}(b)) reveals that the frequency of traveling $f_{tr}$  is 0.0000816.

\subsection{Application of the 0-1 test method for the determination of chaos.}\label{Ap2}

\subsubsection{0-1 test description}

The 0-1 test for chaos was put together for the first time by Georg A. Gottwald and Ian Melbourne \cite{Georg A}. It was designed to distinguish chaotic behavior from the regular behavior of deterministic systems.   When the test result gives 1, or a value very close to 1, the system is chaotic, but when it gives 0 or a value very close to 0, it is regular. It can be summarized as follows:\\
Consider an n-dimensional dynamical system $ \dot{X}=F(X) $ with $X=(x_1,x_2,...,x_n)^T$ a solution of the underlying system also noted as $ X(t) $. If we note by $ \phi(t) $ the observable of the underlying system; the 0-1 test uses the observables to drive the dynamics on well-chosen Euclidean extension and exploit a theorem from Nicol et al. \cite{M. Nicol,P. Ashwin} which explains that the dynamics on the group extension is bounded if the underlying dynamics in nonchaotic, but it behaves like a Brownian motion if dynamics is chaotic. Rather than requiring phase space reconstruction, which is necessary to apply standard Lyapunov exponent methods in the analysis of discrete data, the test works directly with the time series and does not involve any preprocessing of the data \cite{M.T. Rosenstein}. A thorough description and the use of the method can be found in \cite{M.T. Rosenstein,S. Devi,tanekou2020complex,G.A. Gottwald}.

\subsubsection{Application}
\begin{figure}[h!]	
	\includegraphics[width=8.5cm]{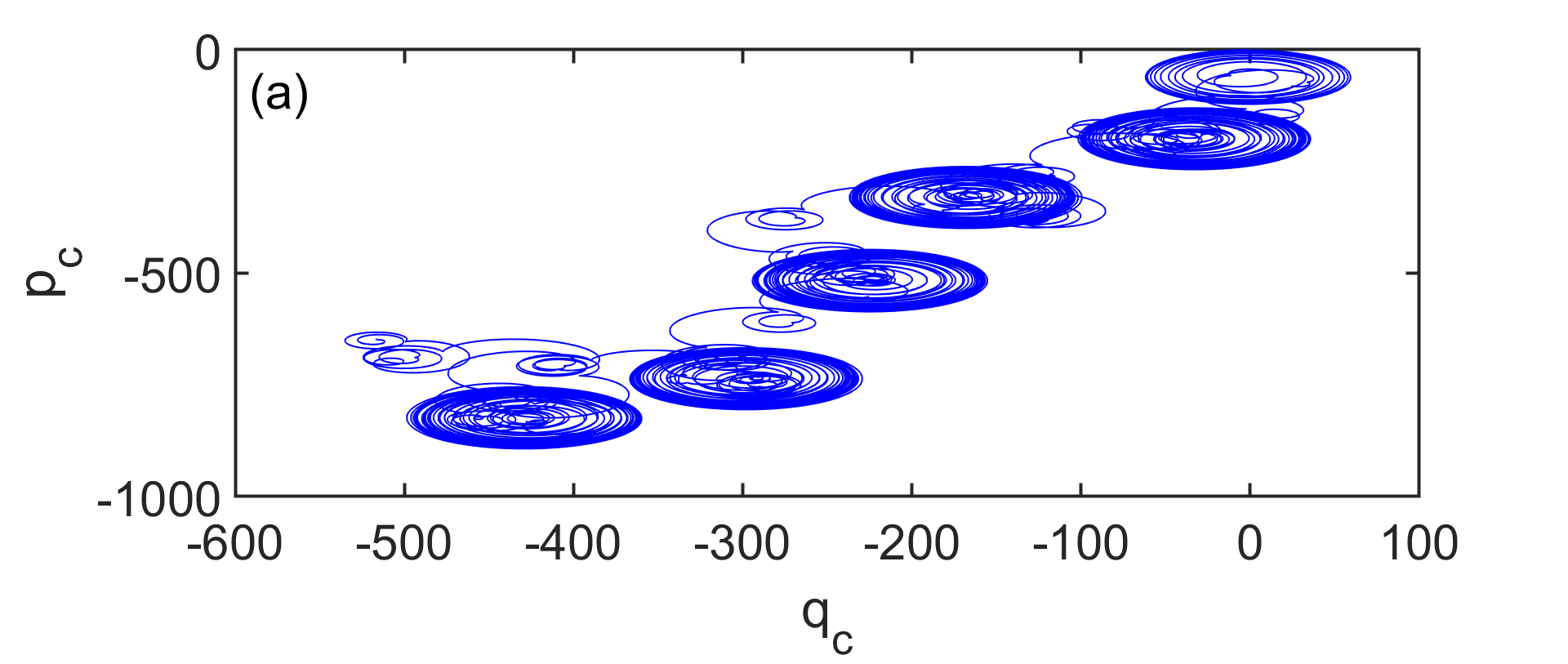}	
	\includegraphics[width=8.5cm]{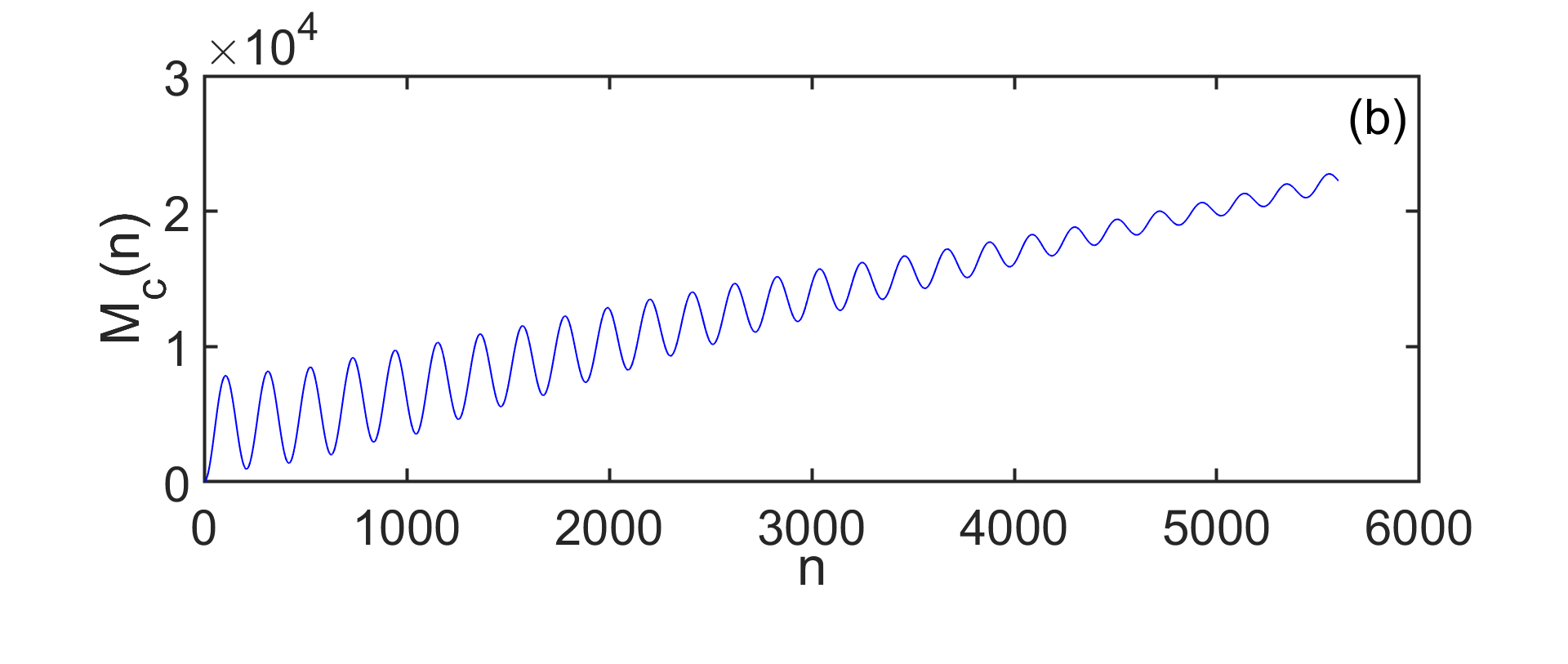}	
	\caption{\label{fig.SGR13} 
		Plot of (a)translation variable $q_c$ versus $p_c$,(b) mean square displacement $M_c$ versus $n$ for the HR model. The trjectories are unbounded.We used
60000 data points and computed $M_c(n)$ for $n=1,...6000$ with $c=0.03$.}
	
\end{figure}

 To confirm that the nodes of the traveling chimera zones of Fig.\ref{fig.SGR10}(a) emit chaotic burstings, we apply to the waveform of the variables x, as presented in red in Fig.\ref{fig.SGR10}(b), the 0-1 test method. It emerges that the trajectories of the phase portrait of the displacement variables, defined as $(p,c)$ by Gottwald and Melbourne, are unbounded(Fig.\ref{fig.SGR13}(a)); then, the oscillations of the mean square displacement evolve asymptotically towards a linear curve (Fig.\ref{fig.SGR13}(b))and finally the value of the correlation coefficient is equal to 0.9906($K_c=0.9906$), i.e., a value very close to 1. All these characteristics constitute proof that the nodes belonging to the traveling zones emit chaotic bursting.

\newpage


\begin{thebibliography}{9}

\bibitem{Purves}
D. Purves, G. J. Augustine, D. Fitzpatrick, W. C. Hall, A.-S. LaMantia, and R. D. Mooney, editors. \textit{Neuroscience}. 6th ed. (Oxford University Press, Sunderland, 2009).

\bibitem{haken}
H. Haken,
\textit{An Introduction to Models and Simulations}
2nd ed. (Springer-Verlag, Berlin, 2008).

\bibitem{izhikevich2007}
E. Izhikevich,
\textit{Dynamical Systems in Neuroscience}
1st ed. (MIT Press, Cambridge, 2006).

\bibitem{pereda}
A. E. Pereda,
Nat. Rev. Neurosci.
\textbf{15} (4), 250 (2014).
	
\bibitem{herrup}
R. W. Williams and K. Herrup,
Annu. Rev. Neurosci \textbf{11} (1), 423 (1988).

\bibitem{swanson}
L. W. Swanson, 
\textit{Brain architecture: understanding the basic plan}
1st ed. (Oxford University Press, New York, 2003).

\bibitem{singer1999}
 W. Singer, Neuron \textbf{24}, 49 (1999).

\bibitem{fries}
P. Fries, Annu. Rev. Neurosci \textbf{32}, 209 (2009).

\bibitem{piko}
A. Pikovsky, J. Kurths and M. Rosenblum,
\textit{Synchronization: A Universal Concept in Nonlinear Sciences}
1st ed. (Cambridge University Press, New York, 2003).

\bibitem{melloni}
L. Melloni, C. Molina, M. Pena, D. Torres, W. Singer and E. Rodriguez, J. Neurosci. \textbf{27}, 2858 (2007).

\bibitem{uhlhaas}
J. Uhlhass and W. Singer, Neuron \textbf{52},155 (2006).


\bibitem{jiruska}
P. Jiruska, M. de Curtis, J. G. Jefferys, C. A. Schevon, S. J. Schiff, and K. Schindler, J. Physiol. \textbf{591}, 787 (2013).

\bibitem{sejnowski}
C. Lainscsek, N. Rungratsameetaweemana, S. S. Cash and T. J. Sejnowski, Chaos \textbf{29}, 121106 (2019).

\bibitem{rosenblum2017}
R. G. Erra, J. L. P. Velazquez, and M. Rosenblum, Front. Comput. Neurosci. \textbf{11}, 98 (2017).

\bibitem{kuramoto2002}
Y. Kuramoto and D. Battogtokh, Nonlinear Phenom. Complex Syst. \textbf{5}, 380 (2002).

\bibitem{abrams2004}
D. M. Abrams and S. H. Strogatz, Phys. Rev. Lett. \textbf{93}, 174102 (2004).

\bibitem{wolfrum2011}
M. Wolfrum and O. E. Omel'chenko, Phys. Rev. E\textbf{84}, 015201(R) (2011).

\bibitem{bick}
C. Bick, M. Sebek and I. Z. Kiss, Phys. Rev. Lett. \textbf{119}, 168301 (2017).

\bibitem{santos2017}
M. S. Santos, J. D. Szezech, F. S. Borges, K. C. Iarosz, I. L. Caldas, A. M. Batista, R. L. Viana, and J. Kurths, 
Chaos Solitons \& Fractals \textbf{101}, 86 (2017).

\bibitem{hindmarshr}
J. Hindmarsh and R. Rose, Proc. R. Soc. B. \textbf{221}, 87 (1984).

\bibitem{andrzejak}
R. G. Andrzejack, C. Rummel, F. Mormann, and K. Schindler, Sci. Rep. \textbf{6}, 23000 (2016).

\bibitem{storace}
M. Storace, D. Linaro, and E. de Lange, Chaos \textbf{18}, 033128 (2008).

\bibitem{majhi}
S. Majhi, B. K. Bera, D. Ghosh and M. Perc, Phys. Life Rev. \textbf{28}, 100 (2019).

\bibitem{hizanidis}
J. Hizanidis, V. G. Kanas, A. Bezerianos, and T. Bountis, Int. J. Bif. \& Chaos \textbf{24}, 1450030 (2014).

\bibitem{bera2016chimera}
B. K. Bera, D. Ghosh and M. Lakshmanan,Phys. Rev. E \textbf{93}, 012205 (2016).

\bibitem{wei2018}
Z. Wei, F. Parastesh, H. Azarnoush, S. Jafari, D. Ghosh, M. Perc and M. Slavinec,Europhys. Lett. \textbf{123}, 48003 (2018).

\bibitem{gluckman96}
B. J. Gluckman, E. J. Neel, T. I. Netoff, W. L. Ditto, M. L. Spano, and S. J. Schiff, J. Neurophysiol. \textbf{76}, 4202 (1996).

\bibitem{gluckman01}
B. J. Gluckman, H. Nguyen, S. L. Weinstein, and S. J. Schiff, J. Neurosci. \textbf{21}, 590 (2001).

\bibitem{park05}
E-H. Park, E. Barreto, B. J. Gluckman, S. J. Schiff, and P. So, J. Comput. Neurosci. \textbf{19}, 53 (2005).


\bibitem{ma2019model}
J. Ma, G. Zhang, T. Hayat, and G. Ren, Nonlinear Dyn.\textbf{95}, 1585 (2019).

\bibitem{Michael V.L.}
M. V. L. Bennett and R. S. Zukin, Neuron \textbf{41}, 495 (2004).

\bibitem{Zahra Shahriari}
Z. Shahriari  , F. Parastesh , M. Jalili , V. Berec, J. Ma  and S. Jafari, EPL \textbf{125}, 60001 (2019).

\bibitem{mishra2017traveling}
A. Mishra, S. Saha, D. Ghosh, G. V. Osipov, and S. K. Dana, Opera Med. Physiol. \textbf{3}, 14 (2017).

\bibitem{graziane}
N. Graziane and Y. Dong, Humana Press \textbf{112}, 157 (2016).

\bibitem{somers}
D. Somers and N. Kopell, Bio. Cybern. \textbf{68}, 393 (1993).

\bibitem{collens}
J. Collens, K. Pusuluri, A. Kelly, D. Knapper, T. Xing, S. Basodi, D. Alacam, and A. L. Shilnikov, Chaos \textbf{30}, 072101 (2020).

\bibitem{travchime}
J. Xie, E. Knobloch, and H-C. Kao,Phys. Rev. E \textbf{90}, 022919 (2014).

\bibitem{yao2015emergence}
N. Yao, Z.-G. Huang, C. Grebogi, and Y-C. Lai, Sci. Rep. \textbf{5}, 12988 (2015).

\bibitem{zhu2012}
Y. Zhu, Y. Li, M. Zhang and J. Yang, EPL  \textbf{97}, 10009 (2012).

\bibitem{laing2009}
C. R. Laing, Physica D \textbf{238}, 1569 (2009).

\bibitem{dai2017two}
Q. Dai, D. Liu, H. Cheng, H. Li, J. Yang, PLOS One \textbf{12}, 0187067 (2017).


\bibitem{omelchenko2011}
M. Wolfrum, O. E. Omel'chenko, S. Yanchuk, and Y. L. Maistrenko, Chaos \textbf{21}, 013112 (2011).

\bibitem{haugland}
S. W. Haugland, L. Schmidt, K. Krischer, Sci. Rep. \textbf{5}, 1 (2015).

\bibitem{gopal2014observation}
R. Gopal, V. K. Chandrasekar, A. Venkatesan, and M. Lakshmanan, Phys. Rev. E \textbf{89}, 052914 (2014).

\bibitem{bera2016imperfect}
B. K. Bera, D. Ghosh, and T. Banerjee, Phys. Rev. E \textbf{94}, 012215 (2016).

\bibitem{dogo}
O. Dogonasheva, D. Kasatkin, B. Gutkin, and D. Zakharov, arXiv:2103.09304v1 [nlin.AO].

\bibitem{kaas}
C. Kaas-Petersen, Plenum Press, New York, NY \textbf{1}, 183 (1987).

\bibitem{hansel}
D. Hansel and H. Sompolinsky, Phys. Rev. Lett. \textbf{68}, 718 (1992).

\bibitem{Hizanidis1}
J. Hizanidis, E. Panagakou, I. Omelchenko, E. Sch{\"o}ll, P. H{\"o}vel, and A. Provata, Phys. Rev. E \textbf{92}, 012915 (2015).

\bibitem{Georg A}
G. A. Gottwald and I. Melbourne, Proc. R. Soc. London, Ser. A \textbf{460}, 603 (2004).

\bibitem{M. Nicol}
M. Nicol, I. Melbourne, and P. Ashwin, Nonlinearity \textbf{14}, 275 (2001).

\bibitem{P. Ashwin}
P. Ashwin, I. Melbourne, and M. Nicol, Physica D \textbf{156}, 364 (2001).

\bibitem{M.T. Rosenstein}
M.T. Rosenstein, J.J. Collins, C.J. De Luca,  Physica D \textbf{65}, 117 (1993).

\bibitem{S. Devi}
S. Devi, S. Singh, A. Sharma, Nonlinear Proc. Geophys.\textbf{20}, 11 (2013).

\bibitem{tanekou2020complex}
G. B. Tanekou, C. F. Fogang, F. B. Pelap, R. Kengne, T.F. Fozin,
B. R. N. Nbendjo, Eur. Phys. J. Plus \textbf{135}, 1 (2020).

\bibitem{G.A. Gottwald}
G. A. Gottwald, I. Melbourne, Nonlinearity \textbf{22}, 1367 (2009).







\end{thebibliography}


\end{document}